\documentclass[11pt]{article}
\usepackage[a4paper,margin=1in]{geometry}
\usepackage{amsmath,amssymb,amsfonts,physics,xcolor}
\usepackage{mathtools}
\usepackage{graphicx}
\usepackage{jheppub}
\usepackage{appendix}
\usepackage{hyperref}
\usepackage{bm}
\DeclareUnicodeCharacter{03B1}{\ensuremath{\alpha}}

\title{Logarithmic corrections to the entropy of near-extremal rotating black holes }

\author[a]{Nabamita Banerjee,}
\author[a]{Koustubh Guha}
\author[b]{and Muktajyoti Saha}

\affiliation[a]{Department of Physics, Indian Institute of Science Education and Research Bhopal,\\ Bhopal Bypass Road, Bhopal MP 462066, India}
\affiliation[b]{International Centre for Theoretical Sciences - TIFR
Bengaluru - 560089, India}
\emailAdd{nabamita@iiserb.ac.in}
\emailAdd{koustubh23@iiserb.ac.in}
\emailAdd{muktajyoti.hepth@gmail.com}

\abstract{
An interesting feature of quantum corrections to the Bekenstein--Hawking entropy is the appearance of logarithmic area terms, whose coefficients are governed by the spectrum of massless fields. For near-extremal black holes, the entropy also receives a logarithmic temperature correction, whose coefficient is determined by the zero modes of the corresponding extremal black holes. In this paper, we study near-extremal rotating black holes in three, four, and five dimensions with an important focus on rotational zero modes. For three-dimensional BTZ black holes, our results are in agreement with the microscopic description. Furthermore, by treating the cosmological constant scale \(\ell\) as an independent parameter, we find an additional \(\log \ell\) correction to the entropy at extremality. For the five-dimensional BMPV black hole, we explicitly construct all the extremal zero modes and use them to evaluate the log temperature corrections to the entropy.
}

\date{}

\begin{document}

\maketitle
%\tableofcontents

\section{Introduction}

Black holes are among the most intriguing objects in general relativity. A suitable identification of their geometrical properties with thermodynamic variables reveals that black holes behave as thermodynamic systems: they possess a temperature and entropy, radiate, and obey the laws of thermodynamics \cite{Bekenstein1973,BardeenCarterHawking1973,Hawking1974}. However, understanding the microscopic origin of black-hole entropy and the mechanism underlying black-hole radiation requires a quantum description of gravity \cite{Hawking1975,GibbonsHawking1977}. The Bekenstein--Hawking area entropy formula,
\begin{equation}
S_{\rm BH}=\frac{A_H}{4G_N},\nonumber
\end{equation}
gives the leading semiclassical entropy of a black hole, while quantum effects generically generate subleading terms of the form
\[
S=S_{\rm BH}+\alpha\log S_{\rm corr}+\mathcal O(1)+\cdots .
\]
The logarithmic correction \cite{Solodukhin1995Conical,Solodukhin1995Nongeometric,Fursaev1995,MannSolodukhin1998,KaulMajumdar2000, Das:2001ic, Carlip:2000nv,Sen:2008vm,BanerjeeGuptaSen2011,Sen:2012kpz,Sen:2012cj,Bhattacharyya:2012wz,Sen2013NonExtremal,JeonLal2017,LiuPandoZayasRatheeZhao2018} is particularly important because it is often the leading controlled quantum correction in the large-area expansion and the coefficient $\alpha$ can be computed entirely from low-energy data i.e. the spectrum of massless fields. This makes the logarithmic term an especially clean infrared probe of the underlying quantum theory. For the zero temperature extremal black holes, this coefficient can be computed from a gravitational path integral on the $AdS_2$ near-horizon geometry -- called the quantum entropy function \cite{Sen:2008vm}. A closely related structure occurs for small temperature near-extremal black holes, where the near-horizon geometry may be treated as a deformation of the extremal throat and the one-loop path integral contains modes that become zero modes in the extremal limit. They produce a $\log T$ correction to the entropy, whose coefficient is again fixed by the spectrum of massless fields. For a generic finite temperature non-extremal black hole, however, there is no analogous decoupled $AdS_2$ throat, and the one-loop calculation generally involves the full Euclidean black-hole geometry, including the appropriate treatment of zero modes and the thermodynamic ensemble.

The main physical importance of the logarithmic correction is that it provides a much sharper test of a microscopic theory of black-hole entropy than the leading area law alone. If the microscopic degeneracy behaves for large charges as
\begin{equation}
d(Q)\sim e^{S_{\rm BH}}S_{\rm corr}^{\alpha}, \nonumber
\end{equation}
then
\[
\log d(Q)=S_{\rm BH}+\alpha\log S_{\rm corr}+\cdots .
\]
Here $S_{\rm corr}$ denotes the correction to the semiclassical entropy and, in general, is a dimensionless combination of black-hole scales, charges, couplings, or moduli. Reproducing $S_{\rm BH}$ tests only the leading exponential growth of the number of states, whereas reproducing the coefficient $\alpha$ probes the subleading power-law prefactor. In supersymmetric extremal black holes, this comparison is particularly powerful because the macroscopic coefficient can be extracted from the low-energy gravitational theory and compared with an independent microscopic state-counting formula. Agreement therefore provides a nontrivial test of string-theoretic or other microscopic descriptions of black holes and constrains candidate ultraviolet completions of gravity. The logarithmic term is sensitive to detailed features of the quantum spectrum and symmetry structure, since one-loop determinants of massless fluctuations, gauge and diffeomorphism zero modes, and the corresponding integration measures can all contribute to its coefficient. It can therefore distinguish theories that reproduce exactly the same classical Bekenstein--Hawking entropy. Consequently, logarithmic corrections provide one of the most precise probes of quantum black-hole physics while testing a proposed quantum theory beyond the universal leading-order area law.

{%Logarithmic corrections also help identify which features of black-hole entropy are universal and which depend on the physical setting, since extremal, near-extremal, and generic non-extremal black holes require conceptually different treatments. Although subleading, $\log S_{\rm corr}$ is parametrically larger than inverse-area corrections for a large black hole and therefore occupies a privileged position in the semiclassical expansion{\color{blue} this statement looks shady: log corrections are imp because they depend only on IR data, shouldn't that be enough motivation?. Consequently, logarithmic corrections provide one of the most precise probes of quantum black-hole physics, connecting semiclassical gravity, quantum fluctuations, and zero-mode effects with microscopic state counting {\color{blue} this part of the sentence doesnt look good},{\color{red}you can modify} while testing a proposed quantum theory beyond the universal leading-order area law. {THIS PARA IS A REPETITION OF THE EARLIER ONES, SO JUST ADDED THE LAST LINE IN PREV. PARA}}

In this paper we concentrate on rotating extremal and near-extremal black holes. The presence of the angular momentum parameter makes their dynamics very rich. In particular, the corresponding euclidean geometries are necessarily complex. This demands considering fluctuations to be complex as well. Hence, while computing their Euclidean gravity partition functions, one often needs to adhere to the Kontsevich-Segal-Witten (KSW) \cite{Kontsevich:2021dmb,Witten:2021nzp,Marolf:2022ntb,Liu:2023jvm} criterion to choose an appropriate contour in the space of complex Euclidean metrics. We shall come back to this point later in section \ref{section5}. 

The most widely studied rotating black holes are the three-dimensional BTZ black holes, which arise in a broad range of contexts in gravitational physics and holography\footnote{There exists no 3-dimensional asymptotically flat black holes.}. The leading quantum corrections to extremal and near-extremal BTZ black holes have been investigated extensively in the literature, although some disagreement remains regarding some parts of the resulting conclusions. For extremal BTZ, the leading quantum correction to the large-entropy degeneracy takes the familiar form
\begin{equation}
S_{\rm ext}=S_0-\frac{3}{2}\log S_0+\cdots ,
\end{equation}
as obtained from Cardy/Rademacher methods and related one-loop analyses~\cite{Carlip:2000nv,Birmingham:2000xd,Govindarajan:2001ee}%{Carlip2000,Birmingham2001,Govindarajan2001}
. This is essentially a large-charge, microcanonical correction and should be distinguished from the low-temperature correction of the near-extremal black hole.  In the latter case, modes that become zero modes at extremality are lifted at small temperature, producing the characteristic Schwarzian scaling \cite{Iliesiu:2020qvm, Iliesiu:2022onk, Banerjee:2023quv},
\begin{equation}
Z_{\rm 1-loop}\sim T^{3/2},
\qquad
S_{\rm near-ext}
=
S_{\rm cl}(T)
+\frac{3}{2}\log\!\left(\frac{T}{T_*}\right)+\cdots .
\end{equation}
For BTZ, this behavior has been reproduced directly in the full geometry using quasinormal-mode methods by Kapec, Law and Toldo~\cite{Kapec:2024zdj} %\cite{KapecLawToldo2025}
, and through the Euclidean fluctuation operator by Kolanowski et al.~\cite{Kolanowski:2024zrq}%~\cite{KolanowskiEtAl2025}
. Recent work suggests that the near-horizon Schwarzian description is not, by itself, the complete quantum answer for BTZ.  Castro, Mancilla and Papadimitriou~\cite{Castro:2025itb} %\cite{CastroMancillaPapadimitriou2025} 
emphasized the nontrivial matching between the asymptotic AdS$_3$ region and the near-horizon effective theory, while Bac, Castro and Jain~\cite{Bac:2026eqj} %\cite{BacCastroJain2026} 
showed that the near-horizon path integral need not coincide with the low-temperature path integral around the full BTZ saddle.  Most recently, Despontin, Detournay, Mancilla and Toldo~\cite{Despontin:2026xzg} %\cite{DespontinEtAl2026} 
analyzed the lifting of Schwarzian and rotational modes explicitly and showed that their contribution depends on the boundary conditions and thermodynamic ensemble.  Thus, the current picture is that the $-\frac{3}{2}\log S_0$ term characterizes the extremal large-charge expansion, whereas the $+\frac{3}{2}\log T$ term is an infrared near-extremal effect associated with soft modes, with the precise full correction controlled by the boundary conditions, ensemble, and the matching of the throat to the asymptotic AdS$_3$ geometry.

For four-dimensional rotating black holes (Kerr, Kerr-Newman and their AdS counterparts), the extremal quantum corrected entropy are well studied. At strict extremality, pure Kerr in four-dimensional Einstein gravity has
$\Delta S=(16/45)\log(A_H/4G_N)+\cdots$~\cite{Sen:2012cj}.
Similarly Extremal Kerr--Newman has a known $\log A_H$ correction whose coefficient depends on the rotation-to-charge ratio in Einstein--Maxwell theory~\cite{Bhattacharyya:2012wz}%BhattacharyyaPandaSen}
; analogous explicit results are known in minimal $\mathcal N=2$ supergravity~\cite{Karan:2019gyn}%KaranBanerjeePanda}
. For the near extremal solutions, there are some open issues regarding the zero modes. The pure near-extremal Kerr has Schwarzian zero modes and their contribution to the entropy ~\cite{Kapec:2024zdj,Rakic:2023vhv} goes as%KapecShetaStromingerToldo,RakicRangamaniTuriaci}
\[
\Delta S=\frac32\log T+\cdots .
\]
The coefficient $3/2$ comes from the tower of normalizable boundary-time reparametrization modes of the extremal NHEK throat, lifted at small finite temperature~\cite{Kapec:2024zdj}%KapecShetaStromingerToldo}
.
 As discussed in \cite{Rakic:2023vhv}%RakicRangamaniTuriaci}
 , the candidate for the rotational zero modes have an important subtlety; the Kerr path-integral analysis explains why they do not simply add a further universal $\frac12\log T$ term in the standard treatment. Additional gauge or rotational nearly-zero modes can occur for charged rotating black holes or in supergravity, and their contribution is theory- and ensemble-dependent~\cite{Modak:2025gvp}%ModakSinghPanda}
 . Hence for these rotating black holes The $\log T$ correction is conceptually distinct from the usual one-loop $\log A_H$ correction.

Let us next briefly summarize the status of five dimensional rotating black holes, the Breckenridge--Myers--Peet--Vafa or BMPV black holes. This still remains a reasonably unexplored system. In the later section we extend upon their construction. The BMPV black hole is the extremal supersymmetric rotating three-charge black hole of five-dimensional supergravity.  In the type-IIB D1--D5--P frame, with D1 charge $Q_1$, D5 charge $Q_5$, momentum $n$, and left angular momentum $J$, its two-derivative entropy is
\begin{equation}
 S_{\rm BH}=2\pi\sqrt{Q_1Q_5n-\frac{J^2}{4}},
\end{equation}
up to the conventional normalization of $J$.  Breckenridge--Myers--Peet--Vafa gave the original microscopic derivation: the weak-coupling D1--D5 system is described by a two-dimensional supersymmetric CFT with effective central charge $c\simeq6Q_1Q_5$, and the asymptotic degeneracy at fixed momentum and $SU(2)_L$ spin reproduces precisely the BMPV area law \cite{Breckenridge:1996is}%BMPV}
. Results are also known beyond the supergravity approximations. Four-derivative terms in five-dimensional $\mathcal N=2$ supergravity, in particular the mixed gauge--gravitational Chern--Simons interaction and its supersymmetric completion, modify both the solution and the Wald entropy.  Castro--Davis--Kraus--Larsen constructed the corresponding corrected spinning black holes and obtained explicit corrections to the BMPV entropy \cite{Castro:2007ci}%CDKL} 
. On the microscopic side, the counting results from the exact D1--D5--P degeneracy have been expanded beyond leading order and found agreement with the Wald entropy of the $R^2$-corrected theory in an overlapping large-charge regime that need not be the Cardy regime \cite{Castro:2008ys}%CM}
. Also in \cite{Banerjee:2008ag}, the author developed the subleading statistical-entropy expansion for BMPV.  These corrections are UV/string-theory dependent and should be distinguished from the infrared logarithmic one-loop corrections.  A recent direct heterotic computation constructs the full asymptotically-flat BMPV solution through first order in $\alpha'$ and obtains the corrected generalized-Wald entropy \cite{Ruiperez:2026rni}.

In the direction of infrared corrections, in \cite{Sen:2012cj} Sen computed the one-loop logarithmic correction to BMPV thermodynamics using the quantum entropy function, i.e. the Euclidean path integral over massless fluctuations in the BMPV near-horizon geometry.  Under the assumption that all the charges get scaled identically,
\[
 Q_1,Q_5,n\sim\Lambda,\qquad J\sim\Lambda^{3/2},
\]
the result he obtained is,
\begin{equation}
 \Delta S_{\log}=-\frac14\,(n_V-3)\log\Lambda,
 \qquad (J\sim\Lambda^{3/2}),
\end{equation}
where $n_V$ is the total number of five-dimensional $U(1)$ gauge fields.  For $J=0$, enhanced rotational symmetry changes the zero-mode contribution and
\begin{equation}
 \Delta S_{\log}=-\frac14\,(n_V+3)\log\Lambda.
\end{equation}
The same coefficients follow from the asymptotic expansion of the microscopic indexed degeneracy. This matching is especially important because logarithmic terms depend only on the low-energy massless spectrum and zero modes, and are insensitive to massive string states and local higher-derivative interactions \cite{Sen:2012cj}%Sen}
. Further progress is made in \cite{Gupta:2019xac}%GMSloc}
, where the relevant five-dimensional localization manifold is determined and subsequently a finite-dimensional localization formula has been proposed for the quantum entropy of BMPV black holes in M-theory on a Calabi--Yau threefold \cite{Gupta:2021roy}%GMS}
. The localized integral has $n_V+2$ variables and its renormalized action is fixed by the two-derivative couplings and the four-derivative gauge--gravitational Chern--Simons coefficients. The known logarithmic correction strongly constrains the one-loop determinant. Thus currently we have quantitative microscopic and macroscopic agreement established for the leading entropy, important higher-derivative terms, and the leading logarithmic quantum correction for the BMPV system.

The present work aims to extend the understanding of the quantum logarithmic corrections for the near-extremal rotating black holes. In this direction, the $\log \, T$ correction for the three dimensional BTZ black holes have recently been analyzed in \cite{Bac:2026eqj}, while some critical comments about the four dimensional Kerr black holes have been presented in \cite{Rakic:2023vhv}. In this paper we have reviewed briefly these works, stressing upon their important physical and technical aspects. We emphasized on points where our understanding differs from the existing literature. Next we present a systematic analysis for the near-extremal entropy corrections for the minimal BMPV solutions. The contributions come from the BMPV near horizon zero modes  the Schwarzian modes, the SU(2) zero modes and the rotational zero modes and the U(1) zero modes. The rotation parameter particularly play a major role in the results. The complete result is given as,
\begin{equation}
S= \frac32\log\left(\frac{T}{T_{\rm Sch.}}\right)+\frac12\log\left(\frac{T}{T_{\rm Rot.}}\right)+\frac32\,\log\left(\frac{T}{T_{ SU(2)}}\right)...
\end{equation}

In the above expression, all the arguments of the logarithms scale as $\log (r_0 T)$, where $r_0$ is the extremal horizon radius and $T$ is the small temperature of the near-extremal black hole. In the typical regime where $r_0$ scales all the BMPV charges uniformly, the heat kernel results of Sen hold \cite{Sen:2012cj}. Temperature $T$ is an independent scale satisfying $r_0 T\ll 1$. This regime was considered in \cite{Banerjee:2023gll} for understanding the near-extremal correction to the entropy. Also, our results show that the rotational zero modes get a special footing as compared to the other two, that we present in section 
\ref{sub-sec:log T corrections near-BMPV}.%{sec5}.

The paper is organized as follows : In section \ref{section2} we give a lightning review of the non-extremal BTZ solution in pure AdS$_3$ gravity. We also discuss the current status of $\log$ corrections to entropy of (near)-extremal BTZ black hole. We present our results, computed from a near-horizon perspective and compare the contributions to $\log T$ corrections due to rotational zero modes with that in~\cite{Bac:2026eqj} and~\cite{Kolanowski:2024zrq}. Sections \ref{section3} and \ref{sec:Kerr rot modes issues} briefly discuss the rotating Kerr black hole solution in asymptotically, flat 4D spacetime and issues like Ricci-flatness in finding the rotational zero modes of the NHEK geometry. In section \ref{section4} we have discussed about the rotating BMPV black hole, a purely bosonic solution in 5D minimal, un-gauged supergravity. This section reviews constructions of the BMPV solution from different perspectives. Subsequently in section \ref{sub-sec:log T corrections near-BMPV}, we present our results for $\log T$ corrections to entropy of near-BMPV black hole. Here, we obtain the near-horizon, near-BMPV geometry from the full non-extremal solution, find the set of zero modes of the kinetic operator evaluated on the near-horizon, BMPV background, and obtain $\log T$ corrections to entropy of near-BMPV solution using known perturbative techniques, purely from a near-horizon perspective. We find that non-trivial contribution to $\log T$ comes only from those zero modes of the extremal background which become slightly non-zero under a small increment in $T$ above extremality. Finally, we conclude with the summary of results, discussions on the same and hint on possible future directions in section \ref{section5}. The appendices contain some simple yet relevant details.

\textbf{Convention:} We use Greek indices $(\mu,\nu,\rho,\dots)$ to denote the full spacetime, while dedicate the lower case Latin indices $(i,j,k,\dots)$ for the $AdS_2$ spacetime. Thus, $g_{\mu\nu}$ denotes components of the full spacetime metric while $x^{i}$ refers to the coordinates $(\theta,\eta)$ on $AdS_2$. The discrete eigenmodes are labeled with an integer index $m$ or $n$. The $SU(2)$ indices are labeled by uppercase Latin letters $(I,J,K,\dots)$. For BMPV, we use the rotation parameters $a$ or $\omega$ related by $a=-\sqrt Q \sin\omega$ interchangeably as and when required. Throughout this paper tensorial objects denoted with a bar are defined with respect to the chosen background.
\section{The BTZ Black Hole}\label{section2}
The three dimensional rotating BTZ black holes are reasonably well studied in the literature. The main references that we refer to are \cite{Brown:1986nw,Banados:1992wn,Carlip:2000nv,Carlip:1996yb,Strominger:1997eq,Banados:1998gg,Govindarajan:2001ee,Birmingham:2000xd,Kolanowski:2024zrq,Azeyanagi:2007qj,Giombi:2008vd,Maloney:2007bna,Ghosh:2019qsp,Poojary:2022roa,Acito:2025lmu}. The important features that we find worth highlighting are :
\begin{itemize}
\item
The Euclidean extremal BTZ geometry is necessarily complex, and no redefinition of its parameters can render it real. Therefore, when studying the Euclidean gravity partition function around such a complex geometry, there is no reason to require the fluctuations to remain real. This observation necessarily justifies the norm adopted in~\cite{Bac:2026eqj} %{Castro}
over that of~\cite{Kolanowski:2024zrq}%{Turiaci}
\footnote{It is also worth mentioning that, in the context of the $AdS_2$ norms defined by Higuchi--Camporesi with respect to the basis functions of~\cite{Camporesi:1994ga}, both the norms defined in~\cite{Bac:2026eqj} %{Castro}
and~\cite{Kolanowski:2024zrq} %{Turiaci}
constitute valid higher-dimensional extensions. However, from the higher-dimensional perspective, the norm defined in~\cite{Bac:2026eqj} %{Castro}
is the relevant one.}. 
\item Zero modes of the extremal kinetic operator are crucial in the logarithmic area corrections to the entropy of extremal BTZ black hole. Furthermore, by treating the cosmological constant, or equivalently the AdS$_3$ radius $\ell$, as a length scale independent from the horizon size, we find novel logarithmic contributions to the extremal entropy.
\item
Following the perturbative technique of~\cite{Banerjee:2023quv}, we compute the contributions of the tensor and rotational zero modes to the near-extremal BTZ entropy. Our results agree with those presented in~\cite{Bac:2026eqj} %{Castro}
 and with the microscopic dual CFT results \cite{Strominger:1997eq} . In particular, the correction to the eigenvalue turns out to vanish for the rotational zero modes. The authors of~\cite{Bac:2026eqj} argue that the vanishing eigenvalue correction signals a breakdown of the perturbative approach. However, we disagree with this conclusion based on their computation. The system under consideration requires a formulation in terms of degenerate perturbation theory. A vanishing first-order correction to the eigenvalue only implies that the degeneracy is not lifted at first order. One must also properly compute the first-order corrections to the fluctuations. In particular, all non-zero modes of the extremal background can, in principle, contribute to these corrections. To the best of our knowledge, the explicit forms of the gauge invariant non zero modes are not yet constructed.  We therefore believe that a more detailed computation is required before drawing any conclusion regarding the validity of the perturbative approach. Furthermore, the compensating gauge transformation required to bring the fluctuations back onto the gauge slice, without changing the physical fluctuations\footnote{We are primarily interested in the physical fluctuations that incorporate the large gauge rotational transformation.}, must always be a small gauge transformation and therefore can never generate a non-normalizable fluctuation. If such a compensating transformation is not possible, that necessarily indicates a failure of the chosen gauge condition, a feature that we shall discuss in the context of Kerr black holes in the next section. It seems highly unlikely that a gauge condition valid for the extremal background would fail at the near-extremal background, with in the framework of perturbation. However it would be good to carry out a thorough computation and we leave it for a future study.
\end{itemize}
We now present the results for the quantum $\log$ corrections to the entropy for extremal and near-extremal BTZ black holes. We begin with the general non-extremal BTZ solution of the Einstein-Hilbert action in pure AdS$_3$ gravity with cosmological constant $\Lambda\,(<0)$, 
\begin{align}
    \begin{split}
        S_{\rm EH}[\,g_{\mu\nu}\,]&=\frac{1}{16\pi G_3}\int d^3x\,\sqrt{-g}\left(R-2\,\Lambda\right)+\text{Bdy. term}
    \end{split}
\end{align}
where, $G_3$ is Newton's constant in three-dimension, $g_{\mu\nu}$ is the 3D metric in mostly positive signature $(-,+,+)$ and $g=\det g_{\mu\nu}$, and $R$ is the corresponding ricci scalar. Particularly for AdS$_3$, $\Lambda=-1/\ell^2$, where, $\ell$ is the AdS$_3$ radius. EOM for the metric is given by,
\begin{align}
    R_{\mu\nu}-\frac{1}{2}\,g_{\mu\nu}\,R+\Lambda g_{\mu\nu}&=0.
\end{align}

The non-extremal BTZ black hole solution paramterized by mass $M$ and angular momentum $J$ is given by,
\begin{align}
    \begin{split}
        ds^2&=-\mathcal{N}(r)^2dt^2+\mathcal{N}(r)^{-2}dr^2+r^2\left(\mathcal{N}^{\varphi}(r)\,dt+d\varphi\right)^2,\label{BTZmetric}
    \end{split}
\end{align}
where, $-\infty\leq t\leq\infty,\,\,\,0\leq r<\infty$, and $0\leq\varphi<2\pi,\,\,\varphi\sim\varphi+2\pi$. The functions
$\mathcal{N}(r)$ and $\mathcal{N}^{\varphi}(r)$ are the lapse and shift functions respectively, and are given by,
\begin{align}
    \begin{split}
        \mathcal{N}(r)=\sqrt{-8G_3M+\frac{r^2}{\ell^2}+\frac{16\,G_3^2\,J^2}{r^2}},&\quad
        \mathcal{N}^{\varphi}(r)=-\frac{4 G_3 J}{r^2}.\label{Metric Functions}
    \end{split}
\end{align}
The horizon radii $r_+$ and $r_-$ can be obtained as positive roots of $\mathcal{N}(r)=0$ given by,
\begin{align}
    \begin{split}
        r_{\pm}&=\ell\left[4G_3\,M\left(1\pm\sqrt{1-\left(\frac{J}{M\ell}\right)^2}\right)\right]^{1/2}.\label{Horizon Radii}
    \end{split}
\end{align}
$|J|\leq M\ell$ ensures the absence of naked singularities. Using (\ref{Horizon Radii}), one can express the mass $M$ and angular momentum $J$ in terms of $r_+$ and $r_-$ as follows,
\begin{align}
    \begin{split}
        M=\frac{r_+^2+r_-^2}{8G_3\ell^2}&,\quad
        J=\frac{r_+r_-}{4G_3\ell}\label{Params-Horizon}.
    \end{split}
\end{align}
The temperature parameter is given by,
 \begin{align}
     \begin{split}
         T_H&=\frac{r_+^2-r_-^2}{2\pi\,\ell^2\,r_+}=\frac{r_+^4-16G_3^2\,J^2\,\ell^2}{2\pi\,\ell^2\,r_+^3}=\frac{r_+^4-r_0^4}{2\pi\,\ell^2\,r_+^3}.\label{Hawking Temp}
     \end{split}
 \end{align}

 In the extremal case, i.e., when $|J|=M\ell$, the two horizons coalesce at $r_+=r_-=r_0$,
 \begin{align}
     \begin{split}
     r_0&=\left.2\ell\sqrt{G_3M}\,\right|_{|J|=M\ell}=2\sqrt{G_3|J|\ell}\label{Extremal Horizon}.
     \end{split}
 \end{align}
We will take the extremal limit by first considering a small temperature parameter ($ T\to 0$) keeping the angular momentum to be fixed at its extremal value $4G_3 J \ell = r_0^2$. This will allow us to write down the near-horizon near-extremal geometry. In this limit, the parameters of the black hole have the following expansion,
\begin{align}
    r_+&=r_0\left[1+\frac{\ell^2\,\pi}{2 \,r_0}\,T+\frac{3\,\ell^4\,\pi^2}{8\,r_0^2}\,T^2\right]+\mathcal{O}(T^3), \\[5pt]
    M&=\frac {r_ 0^2} {4 G_3 \ell^2}\left[1 + \frac {\pi^2 \ell^4} {2 r_ 0^2} T^2 + \frac {\pi^3 \ell^6}{2 r_ 0^3} T^3 \right] + \mathcal {O} (T^4).
\end{align}
Now, to zoom near the horizon, we perform the following transformation of coordinates from $(t,r,\varphi)$ to $(\theta,\eta,\phi)$,
 \begin{align}
     \begin{split}
         t&%=\frac{\theta}{2\,\pi\,T}
         =\frac{\theta\,\ell^2}{2\,\pi\,r_0\,\epsilon}
         , \quad
         r%=r_++\frac{\ell^2\,\pi}{2}\,T\left(\cosh{\eta}-1\right)
         =r_++\frac{\pi\,r_0}{2}\epsilon\left(\cosh\eta-1\right)
         , \quad
         \varphi%=\phi+\frac{1}{2\,\pi\,\ell\,T}\,\theta-\frac{\ell}{2\,r_0}\,\theta
         =\phi-\frac{\theta\,\ell}{2\,r_0}\left(1-\frac{1}{\pi\epsilon}\right),
         \label{near-horizon coordinate transformation}
     \end{split}
 \end{align}
 Here, the near-extremal perturbative parameter is $\epsilon\equiv\frac{\ell^2 T}{r_0}\ll 1$. At $\epsilon\to 0$ limit, we obtain the extremal $AdS_2$ geometry fibered over $\rm S^1$,
 \begin{align}
     ds_0^2 = \frac{\ell^2}{4}(\sinh^2\eta \,d\theta^2 + d\eta^2) + r_0^2\left(d\phi -i\, \frac{\ell}{r_0}\sinh^2 \left(\frac{\eta}{2}\right)d\theta \right)^2. \label{ext-NH-BTZ}
 \end{align}
where, $\theta\in[0,2\pi),\,\theta\sim\theta+2\pi$, $\eta\in[0,\infty)$, and $\phi\in[0,2\pi),\,\phi\sim\phi+2\pi$. The first order correction at $\mathcal{O}(\epsilon)$ is given by,
\begin{align}
    ds_c^2 &=\frac{\pi \ell^2}{8}\left(2+\cosh{\eta}\right)\tanh^2\left(\frac{\eta}{2}\right)\left(\sinh^2\eta\,d\theta^2 + d\eta^2\right) + \pi r_0^2\,\cosh\eta \left(d\phi-i\, \frac{\ell}{r_0}\sinh^2 \left(\frac{\eta}{2}\right)d\theta \right)^2 \nonumber\\
    & -i\,\frac{\pi\ell\, r_0}{4}\sinh^2\left(\frac{\eta}{2}\right)\left(-11+\cosh\eta\right)\,d\theta d\phi+\pi\ell^2\sinh^6 \left(\frac{\eta}{2}\right)d\theta^2.
\end{align}

\subsection{Logarithmic corrections to extremal entropy}
The quadratic action in Euclidean signature governing the massless fluctuations at one-loop order is given by,
\begin{align}\label{quad-action}
    \begin{split}
        \delta^2S_{\rm EH}\,[h_{\mu\nu}]&=-\frac{1}{16\pi G_3}\int d^3x\,\sqrt{\bar g}\,\,\frac{1}{4}\,h_{\mu\nu}\bigg[\,\bar g^{\mu\rho}\,\bar g^{\nu\sigma}\big(\bar\Box+\frac{2}{\ell^2}\big)-\frac{1}{2}\,\bar g^{\mu\nu}\,\bar g^{\rho\sigma}\,\bar\Box\,\bigg]h_{\rho\sigma}.
    \end{split}
\end{align}
Here $\bar \Box= \bar \nabla _\mu \bar \nabla ^\mu$ and the bar denotes a quantity evaluated on a chosen background. To understand the logarithmic corrections to the entropy of the extremal BTZ solution, we compute the Euclidean gravity path integral on the geometry \eqref{ext-NH-BTZ} at one-loop order. We consider the path integral of fluctuations with action \eqref{quad-action} on the extremal near-horizon geometry. The gravitational fluctuations on this geometry can be expanded into a complete basis on $AdS_2$ and $\rm S^1$, and they can be separated into nonzero and zero modes. The nonzero mode eigenvalues scale as $\frac{1}{\ell^2}$, when $r_0 \gg \ell$, while they scale as $\frac{1}{r_0^2}$ when $r_0 \ll \ell$. The logarithmic contributions coming from these nonzero modes in either regime become trivial as the conformal anomaly is zero in three dimensions \cite{Sen2013NonExtremal}. Let us denote the scaling of the nonzero eigenvalues collectively as $\frac{1}{a^2}$, where $a$ can be either $r_0$ or $\ell$ in the regimes $r_0 \ll \ell$ and $r_0 \gg \ell$ respectively. Then, the nonzero modes contribute as,
\begin{align}\label{nonzero-BTZ}
    \log Z_{\text{nz}} \sim - \mathcal{N}_{\rm zm}\log a,
\end{align}
Here $\mathcal{N}_{\rm zm}$ is the number of zero modes.

Other non-trivialities appear from the zero mode fluctuations. The definition and construction of zero modes have been presented in details in section \ref{section4}. For the BTZ black hole, these have already been constructed in \cite{Kolanowski:2024zrq} and \cite{Bac:2026eqj}. Here we only state the results. Physically, the zero modes are associated with the large diffeomorphisms of the two-dimensional throat and large gauge transformations near the $AdS_2$ boundary. These two types of zero modes are dubbed as tensor or Schwarzian modes and rotational zero modes. Their contribution to the path integral can be fixed from their normalization condition,
\begin{align}
    \begin{split}
        \int\,[\mathcal{D}h_{\mu\nu}]\exp\left[-\int\,d^3x\,\sqrt{\bar g}\,\,\bar g^{\mu\nu}\,\bar g^{\rho\sigma}\,h_{\mu\rho}\,h_{\nu\sigma}\right]&=1.\label{Norm_Cond}
    \end{split}
\end{align}
The prescription for finding the contribution to the logarithmic correction of black hole entropy from the metric zero modes is to extract the dependence on the horizon size from the background metric and suitably normalize the path integral measure in (\ref{Norm_Cond}). Additional contributions may arise from the Jacobian associated with the change of variables from the metric zero modes to the local, non-normalizable parameters (independent of the horizon size) that generate such fluctuations. Now, following the above prescription if the $i^{\text{th}}$ zero mode contributes a factor of $a^{\beta_i}$ ($a$ being the relevant length scale) and if there are $\mathcal{N}_{\rm zm}^{i}$ number of such modes, then the overall contribution to the partition function from all the zero modes in the spectrum is given as,
\begin{align}
    \begin{split}
        Z&=a^{\,\sum\limits_{i}\,\beta_i\,\mathcal{N}_{\rm zm}^{i}}.\label{PF_ZM_Contri}
    \end{split}
\end{align}
From \eqref{ext-NH-BTZ}, we observe the following scalings of the near-horizon, extremal BTZ solution, $g_{ij}\sim\ell^2\,r_0^0\,,\,\,g_{i\phi}\sim\ell\,r_0\,,\,\,g_{\phi\phi}\sim\ell^0\,r_0^2$. While, $g^{ij}\sim\ell^{-2}\,r_0^0\,,\,\,g^{i\phi}\sim\ell^{-1}\,r_0^{-1}\,,\,\,g^{\phi\phi}\sim\ell^0\,r_0^{-2}$. %Note that $g^{\mu\nu}$ is independent of the radius of the extremal horizon $r_0$, $g^{\mu\phi}\propto r_0^{-1}$, and $g^{\phi\phi}\propto r_0^{-2}$ (where $x^{\mu}=\{\theta,\eta\}$). 
Also, $\sqrt{ g}=\frac{\ell^2}{4}r_0\sinh{\eta}\,\,\Rightarrow\,\sqrt{ g}\sim \ell^2\,r_0$. Using these scalings, we will now find the contributions from two different types of zero modes.

\subsubsection*{Schwarzian zero modes}
We need to extract the $r_0,\ell$-dependence from the terms associated with the near-horizon, extremal background metric in the exponent in the definition of the normalization condition in (\ref{Norm_Cond}) and suitably adjust the path integral measure. First, we do this for Schwarzian or tensor modes\footnote{The tensors on $AdS_2$ have been constructed in details in \cite{Camporesi:1994ga}.}, which have a $\text{tensor}_{\rm AdS_2}\otimes\text{scalar}_{\rm S^1}$ structure given by,
\begin{align}
    \begin{split}
        &h^{(n)}_{\mu\nu}{}^{\rm Sch.}dx^\mu\,dx^\nu=\frac{\ell}{2\sqrt{2}\,\pi\sqrt{r_0}}\left[\frac{|n|(n^2-1)}{2}\right]^{1/2}\,\frac{(\sinh{\eta})^{|n|-2}}{(1+\cosh{\eta})^{|n|}}\,e^{in\theta}\\[5pt]
        &\quad\times\left(d\eta^2+2i\,\text{sgn}(n)\,\sinh{\eta}\,d\eta\,d\theta-\sinh^2{\eta}\,d\theta^2\right),\quad n\in\mathbb{Z},\,|n|\geq2.  \\[5pt]%\label{TZM_Explicit}
    \end{split}
\end{align}
They are normalized according to the following orthonormality relation,
\begin{align}
    \begin{split}
        \langle f_{n}|f_{m}\rangle&=\delta_{-n,m},
    \end{split}
\end{align}
where, $|f_n\rangle$ denotes the state corresponding to the mode with label $n$ defined above. The inner product is defined as \cite{Bac:2026eqj},
\begin{align}\label{inner-prod}
    \begin{split}
        \langle f_{n}|f_{m}\rangle& \equiv\int d^3x\,\sqrt{\bar g}\,f^{\mu\nu}_n(x)\,f_{m}{}_{\mu\nu}(x).
    \end{split}
\end{align}
The norm defined above gains special importance in the case of rotational modes. For the tensor modes, however, one can easily check that the same boils down to the usual orthonormality relation defined on a $\mathbb{C}$-valued Hilbert space\footnote{This definition is usually used for real backgrounds,
\begin{align*}
 ( f_{n}|f_{m})&=\delta_{n,m}, \qquad
           ( f_n|f_m)\equiv\int d^3x\,\sqrt{\bar{g}}\,f^{*\mu\nu}_n(x)\,f_{m}{}_{\mu\nu}(x). 
           \end{align*}
           }. 
           
It is interesting to note that,
\begin{align}
    \begin{split}
        \bar g^{\mu\nu}\,h^{(n)}_{\mu\nu}{}^{\rm Sch.}&=0\,,\quad \bar \nabla_\rho \bar \nabla^\rho\,h^{(n)}_{\mu\nu}{}^{\rm Sch.}=-\dfrac{2}{\ell^2}\,h^{(n)}_{\mu\nu}{}^{\rm Sch.},
    \end{split}
\end{align}
from which it is evident that these modes have zero extremal action at quadratic order. As we shall later see, these modes reproduce the log $T$ corrections that can also be mapped to be originated from an effective 1D Schwarzian theory.

To understand the scaling behaviours, we expand the contracted terms in the exponent for the tensor zero mode fluctuations as follows,
\begin{align}
    \begin{split}
        \bar g^{\mu\nu}\,\bar g^{\rho\sigma}\,h_{\mu\rho}\,h_{\nu\sigma}&=\bar g^{ij}\,\bar g^{kl}\,h_{ik}\,h_{jl},\qquad h_{\mu\nu}=\sum_n\,c_n h_{\mu\nu}^{(n)}{}^{\rm Sch.}.
    \end{split}
\end{align}
Here, $c_n$ scales as $L^{3/2}$, where $L$ is some length parameter. For tensor zero mode fluctuations we have, $h_{i\phi}=h_{\phi i}=0=h_{\phi\phi}$. For the background metric,  $\bar g^{ij}\sim \ell^{-2}, \sqrt{\bar g} \sim r_0\,\ell^2$. Thus, the normalization condition for the tensor zero modes may re-written as,
\begin{align}
    \begin{split}
        \int\,[\mathcal{D}h_{\mu\nu}]\exp\left[-\,r_0\,\ell^{-2}\int\,d^3x\,\sqrt{\bar g}{}^{\,0}\,\bar g^0{}^{\mu\nu}\,\bar g^0{}^{\rho\sigma}\,h_{\mu\rho}\,h_{\nu\sigma}\right]&=1,\label{Norm_Cond_TZM}
    \end{split}
\end{align}where, $\bar{g}^{0}$ is parameter-independent. Hence, from (\ref{Norm_Cond_TZM}) it can be seen the correctly normalized path integration measure is $\left[\mathcal{D}h_{\mu\nu}\right]=\prod_{x,(\mu\nu)}\,\sqrt{r_0}\,\ell^{-1}\,dh_{\mu\nu}(x)$. Now, the tensor zero modes are generated by non-normalizable diffeomorphism parameters as shown below,
\begin{align}
    \begin{split}
        h_{\mu\nu}&= \bar\nabla_{\mu}\xi_{\nu}+\bar\nabla_{\nu}\xi_{\mu},
    \end{split}
\end{align}where, $\xi^\mu$ is the diffeomorphism parameter, whose integration limits are independent of the length scales. We make a change of variables from the fluctuations to the diffeomorphism generators. Analyzing the background and connection pieces, we find that the resulting Jacobian contributes a factor of $\ell^{2}\,r_0^0$ per zero mode. Thus, for each tensor zero mode, we get a total factor of $\sqrt{r_0}\ell$, giving us the following contribution to the logarithm of the partition function,
\begin{align}\label{tzm-BTZ}
    \begin{split}
        \log\,Z&\sim \frac{1}{2}\mathcal{N}^{\text{tzm}}\,\log\,r_0+\mathcal{N}^{\text{tzm}}\,\log\,\ell, 
    \end{split}
\end{align}
where, the number of such zero modes $\mathcal{N}^{\rm tzm}$ is given by\footnote{We have regulated the divergence coming from infinite length of the $AdS_2$ throat to extract the finite contribution to the number of zero modes in \eqref{tzm-num} and \eqref{rzm-num}.}, 
\begin{align}
    \begin{split}\label{tzm-num}
        \mathcal{N}^{\rm tzm}&=\int\,d^3x\,\sqrt{\bar{g}}\,\bar{K}^{\rm tzm}=3\left(\cosh{\eta_0}-1\right) \xrightarrow[]{\rm reg} -3,
    \end{split}
\end{align}
with,
\begin{align}
    \begin{split}
        \bar{K}^{\rm tzm}(x)&=\sum_{n}\,\bar{g}^{\mu\nu}\,\bar{g}^{\rho\sigma}\,h_{\mu\rho}^{(-n)}{}^{\rm Sch.}\,h^{(n)}_{\nu\sigma}{}^{\rm Sch.}=\dfrac{3}{\pi^2\,\ell^2\,r_0},
    \end{split}
\end{align}
being the corresponding heat kernel trace. Since the volume of $AdS_2$ is infinite, we regulate the same by introducing a radial cut-off $\eta_0$, the upper limit of the integration over the radial coordinate $\eta$.  

\subsubsection*{Rotational Zero Modes}

The non-normalizable vector field that generates the rotational zero modes constructed explicitly in \cite{Kolanowski:2024zrq} has the following form,
\begin{align}
    \begin{split}
        \xi_n&=H_n(x^{i})\,\partial_{\phi}-\,\text{sgn}(n)\frac{r_0\,\ell}{2}\,\nabla^{\mu}H_n(x^{i})\,\partial_{\mu},\label{Vec_Field_RZM}
    \end{split}
\end{align}
where,
\begin{align}\label{eq:scalar_AdS2}
     H_{n}(x^{i})&=\frac{1}{\sqrt{2\pi|n|}}\,\left(\frac{\sinh{\eta}}{1+\cosh{\eta}}\right)^{|n|}\,e^{in\theta}\,,\quad n\in\mathbb{Z},\,|n|\geq1.
\end{align}
is a non-normalizable scalar function on $AdS_2$, independent of our length scales. Here, the physical data about the fluctuations is encoded in the first term of $ \xi$, whereas the second piece is a small compensating gauge transformation needed to bring back the generated metric fluctuations to the chosen gauge slice\footnote{We have chosen harmonic gauge for metric. It can be easily verified that the norm of the compensating gauge transformation parameter vanishes at the boundary, justifying it as a small gauge.}. The rotational modes are associated with pure diffeomorphisms generated by $\{\xi_n\}$ and are given by,
\begin{align}
    h^{(n)}_{\mu\nu}{}^{\rm Rot.}&=2\,\bar\nabla_{(\mu}\,\xi_n\,{}_{\nu)}.
\end{align}
Again, these modes are normalized with respect to the inner product \eqref{inner-prod}. The corresponding fluctuations generated are as follows,
\begin{align}
    \begin{split}
        h_{\mu\nu}&=\sum_{|n|\geq1}c_nh^{(n)}_{\mu\nu}{}^{\rm Rot.}=\bar\nabla_{\mu}\,\xi_{\nu}+\bar\nabla_{\nu}\,\xi_{\mu},\qquad\xi=\sum_{|n|\geq1}c_n\,\xi_n.\label{diffeo_RZM}
    \end{split}
\end{align}
For the fluctuations constructed above the integrand in the exponent of the path integral normalization condition in (\ref{Norm_Cond}) scales as,
\begin{align}
    \begin{split}
        \sqrt{\bar g}\,\bar g^{\mu\nu}\,\bar g^{\rho\sigma}\,h_{\mu\rho}\,h_{\nu\sigma}\,&\sim\,\ell^{0}\,r_0^3,\label{Overall_rh_scaling_TZM}
    \end{split}
\end{align}where, in obtaining the scale dependence in (\ref{Overall_rh_scaling_TZM}) we have implicitly changed the path integration variables from the metric fluctuations to the parameters generating them, whose integration range is independent of the length parameters. Thus, each rotational zero mode contributes a factor of $r_0^{3/2}$ via the measure. Hence, the overall contribution from the rotational zero modes to the logarithm of the partition function may be given as,
\begin{align}\label{rzm-BTZ}
    \begin{split}
        \log Z&\sim\,\frac{3}{2}\,\mathcal{N}^{\text{rzm}}\,\log\,r_0,
    \end{split}
\end{align}where, $\mathcal{N}^{\text{rzm}} = -1$ denotes the number of rotational zero modes. It is defined from the heat kernel trace as follows,
\begin{align}\label{rzm-num}
    \bar{K}^{\rm rzm}(x)&=\dfrac{1}{\pi^2\,\ell^2\,r_0}\,,\quad\mathcal{N}^{\rm rzm}=\int\,d^3x\,\sqrt{\bar g}\,\bar{K}^{\rm rzm}
        =\cosh{\eta_0}-1 \xrightarrow[]{\rm reg} -1.
\end{align}

\subsubsection*{Complete results}
Now we combine all the contributions from nonzero, tensor and rotational zero modes \eqref{nonzero-BTZ}, \eqref{tzm-BTZ} and \eqref{rzm-BTZ}, and find the total contributions in three regimes,
\begin{enumerate}
    \item \textbf{$r_0 \gg \ell$:} 
    \begin{align}\label{BTZER1}
        \log Z &\sim \left(\frac{1}{2}\mathcal{N}^{\text{tzm}} + \frac{3}{2}\,\mathcal{N}^{\text{rzm}}\right)\,\log\,r_0+\mathcal{N}^{\text{tzm}}\,\log\,\ell - \left(\mathcal{N}^{\text{tzm}} + \mathcal{N}^{\text{rzm}}\right)\log \ell ,\nonumber \\
        &\sim -3\log r_0 + \log \ell
    \end{align}
    \item \textbf{$r_0 \ll \ell$:} 
    \begin{align}\label{BTZER2}
        \log Z &\sim \left(\frac{1}{2}\mathcal{N}^{\text{tzm}} + \frac{3}{2}\,\mathcal{N}^{\text{rzm}}\right)\,\log\,r_0+\mathcal{N}^{\text{tzm}}\,\log\,\ell - \left(\mathcal{N}^{\text{tzm}} + \mathcal{N}^{\text{rzm}}\right)\log r_0 ,\nonumber \\
        &\sim \log r_0 - 3\log \ell
    \end{align}
    \item \textbf{$r_0 \sim \ell \sim a$:}
    \begin{align}\label{BTZER3}
        \log Z \sim -2\log a
    \end{align}
\end{enumerate}
Next, we will consider the logarithmic contributions for near-extremal BTZ entropy.

\subsection {Logarithmic corrections to near-extremal entropy}

The prescription for computing the $\log \, T$ correction was formulated in an earlier work \cite{Banerjee:2023quv}. The prescription is to treat the near-extremal geometry as a small-temperature deformation of the extremal near-horizon geometry, rather than taking the $T\to 0$ limit of a generic non-extremal one-loop result. One writes the quadratic fluctuation operator schematically as
\[
\Delta=\Delta_0+T\,\Delta^{(1)}+\cdots,
\]
where $\Delta_0$ is the kinetic operator on the extremal near-horizon background. Since the eigenfunctions of $\Delta_0$ are known, ordinary first-order perturbation theory can be used to determine how its eigenvalues shift at small $T$. The crucial observation is that only the exact zero modes of the extremal operator can generate a $\log \, T$ term: if $\lambda_n^{(0)}=0$, then generically
\[
\lambda_n \sim T\,\lambda_n^{(1)}.
\]
The correction piece $\lambda_n^{(1)}$ is computed from the quadratic action around the extremal background, as we present in the later sections. Thus the one-loop determinant contains
\[
\log \lambda_n \sim \log T.
\]
By contrast, modes with nonzero extremal eigenvalues give only regular corrections in $T$, while modes that remain exact zero modes of the near-extremal operator must be treated separately through their zero-mode measure and they will never contribute a pure $\log \, T$ term. 

Operationally, the prescription is therefore to identify all normalizable zero modes of the extremal near-horizon kinetic operator, evaluate the matrix elements of the $O(T)$ correction $\Delta^{(1)}$ within this zero-mode subspace, and use the resulting lifted eigenvalues in the one-loop determinant. The coefficient of $\log T$ is then determined by the number and degeneracies of the zero modes that are lifted linearly in $T$. 
Thus, the near-extremal $\log T$ correction is controlled by extremal zero modes that become parametrically light when extremality is weakly broken. We adopt this prescription in the present work. 
\\
\subsubsection*{log T corrections from Schwarzian modes }

The near-extremal corrected  eigenvalue corresponding to the $n^{\text{th}}$ Schwarzian mode is found to be,
\begin{align}
    \begin{split}
        \lambda^{\rm Sch.}_n&=\frac {|n|\,\pi\,\epsilon} {256\,\ell^2}\bigg[\cosh {4\eta _ 0} + 
    4 (|n| + 1)\left (8 n^2 - 1 \right)\cosh {\eta_ 0} + 
    4\,(2|n|(|n| + 2) + 1)\cosh {2\eta _ 0}\\[5pt]
    &\quad\quad\quad\quad\quad+ 
    4 (|n| + 1)\cosh {3\eta _ 0} + 64|n|^3+88n^2-48|n|%8 n (n (8 n + 11) - 6) 
    - 69 \bigg]\\[5pt]
    &\quad\quad\quad\quad\quad\times\left (\coth {\eta_ 0} + \csch {\eta_ 0} \right)^{-2 |n|}\csch^2\left (\frac {\eta_ 0} {2} \right)\sech^6\left(\frac {\eta_ 0} {2} \right).
    \end{split}
\end{align}
In the asymptotic limit of radial cut-off $\eta_0$, the corrected eigenvalue evaluates to,
\begin{align}
    \begin{split}
        \lambda^{\rm Sch.}_n&=\frac {|n|\,\pi\,\epsilon} {2\,\ell^2}=\frac {|n|\,\pi\, T} {2\, r_ 0}.
    \end{split}
\end{align}
The corrected quadratic action is as follows,
\begin{align}
    \begin{split}
        S^{(2)}&\sim\sum_{|n|\geq2}c_{-n}\,c_n\times\frac1{16\pi G_3}\times\lambda^{\rm Sch.}_n\\[5pt]
        &\sim\sum_{|n|\geq2}32\,G_3\ell^2\times\frac{\tilde c_{-n}\,\tilde c_n}{16\pi G_3}\times\lambda^{\rm Sch.}_n,\qquad\tilde c_{\pm n}\in\mathbb{C}.
    \end{split}
\end{align}
Hence, the logarithm of the partition function due to the Schwarzian modes is given by,
\begin{align}
    \begin{split}
        \log Z&\sim-\frac{1}{2}\sum_{|n|\geq2}\,\log\left(\frac{|n|\,\ell^2\,T}{r_0}\right)\sim-2\cdot\frac12\,\sum_{n\geq2}\,\log\left(\frac{n\,\ell^2\,T}{r_0}\right).
    \end{split}
\end{align}
Zeta-regularizing\footnote{Define $\mathcal{Z}_\alpha(s)=\alpha^s\zeta(ks)$, where, $\zeta(s)=\sum_{n=1}^{\infty}n^{-s}$ is the zeta function. Then one can show that $\sum_{n=2}^{\infty}\log \left(\alpha\,n^k\right)=(\zeta(0)-1)\log\alpha-k\cdot\zeta'(0)$, where, $\zeta(0)=-\frac12$, and $\zeta'(0)=-\frac12\,\log2\pi$.} the above sum we get,
\begin{align}\label{BTZNER}
    \begin{split}
        \log Z^{\rm reg.}&=\frac{3}{2}\,\log\left(\frac{T}{T_{\rm Sch.}}\right),\qquad T_{\rm Sch.}=\frac{r_0}{\ell^2}
    \end{split}
\end{align}
\subsubsection*{ $\log T$ corrections from rotational zero modes }
The expression for the corrected eigenvalues associated with the rotational zero modes are exponentially damped in $\eta_0$, the radial cut-off, once we consider $\eta_0$ to be large,
\begin{align}
    \lambda^{(c)}_n{}^{\rm Rot.}\xrightarrow[]{\,\,\eta_0\rightarrow\infty\,\,}\dfrac {2\,\pi\, n^2\,T} {r_0}\,e^{-\eta_0}\sim0.
\end{align}
Thus, the corrected eigenvalues\footnote{However, if continued with the usual Hilbert space definition of orthonormality, one would end up with an eigenvalue correction that diverges linearly in $\eta_0$ for every rotational mode described by the label $n$.} are essentially zero and hence, the rotational modes do not contribute to $\log T$ corrections. Since, these modes continue to be zero modes of the kinetic operator in the near-extremal background, they do not encounter an uplift unlike the tensor zero modes which become slightly non-zero in the near-extremal background. The result is consistent with the dual CFT results and thus justifies the near-horizon computation. To the best of our understanding there is a dissent in the literature regarding the rotational zero modes and in the prescription to incorporate their effects on the near-extremal entropy. According to~\cite{Kolanowski:2024zrq}%{Turiaci}
, these modes are considered to be not present and according to~\cite{Bac:2026eqj}%{castro}
, they can not be treated perturbatively around the extremal BTZ. We think that it requires involved computations to come to any of the two conclusions. In fact, with present results, we find that with appropriate complex fluctuations and degenerate perturbation technique, they do reproduce the right CFT result. This completes the review on the status of quantum corrections to the entropy of extremal and near-extremal BTZ black holes. 

\section{The Kerr Black Hole}\label{section3}

The four dimensional rotating black holes are Kerr, Kerr-Newman black holes in asymptotically flat space-times and their (A)dS cousins. All these black holes have an extremal counterpart and possible near-extremal extensions. These systems are reasonably studied in \cite{Sen:2012cj, Bhattacharyya:2012wz,PathakPorfyriadisStromingerVarela2017,Karan:2019gyn,KaranPanda2021,KaranPanda2021Generalized,David:2021eoq,DavidGonzalezLezcanoNianPandoZayas2022,Kapec:2023ruw,Rakic:2023vhv,MaulikPandoZayasRayZhang2024,Kolanowski:2024zrq,Arnaudo:2024bbd,KaranPuniaBiswas2025,MarianiToldo2026,Modak:2025gvp,MaulikMengPandoZayas2026,LuoPandoZayas2026}. Among these the ordinary Kerr still has some interesting open issues, that we point out in this section. 

The Euclidean near-horizon extremal Kerr (NHEK) metric can be written as
\[
 ds^2
 =
 2J\,\Gamma(\theta)
 \left[
 (r^2-1)\,d\tau^2+\frac{dr^2}{r^2-1}+d\theta^2
 +\hat\Lambda^2(\theta)(d\phi+i(r-1)\,d\tau)^2
 \right]
\]
with
\[
\Gamma(\theta)=\frac{1+\cos^2\theta}{2},
\qquad
\hat\Lambda(\theta)=\frac{2\sin\theta}{1+\cos^2\theta}.
\]
It is useful to note the overall $\theta$-dependent factor:
\[
\bar g_{\mu\nu}
=
\Omega^2(\theta)\,\hat g_{\mu\nu},
\qquad
\Omega^2(\theta)=2J\Gamma(\theta)=J(1+\cos^2\theta).
\]
The conformal factor is part of the physical background metric and it is not a Weyl gauge freedom of pure Einstein gravity. The semiclassical thermodynamics and its leading quantum corrections are well studied in the literature from 4-dimensional perspective \cite{Bhattacharyya:2012wz} and also from a dimensionally reduced 2-dimensional perspective \cite{Sen:2012cj}. The log corrections to the semiclassical entropy that from the extremal zero modes, clearly indicates the existence of three tensor zero modes, associated with the $SL(2,\mathbb{R})$ isometries and one rotational zero modes associated with the $U(1)$ axial symmetry. While the 4-dimensional tensor zero modes have been constructed, the rotational zero modes are still not found \cite{Rakic:2023vhv,Kolanowski:2024zrq}. Below we lay out the issues related to the rotational zero modes.

\subsection{Issues with rotational zero modes}\label{sec:Kerr rot modes issues}

A natural first choice for the gauge fixing condition that we begin with is the full background-covariant harmonic gauge condition, i,e,
\[
\mathcal F_\nu[h]\equiv\bar\nabla^\mu h_{\mu\nu}
-\frac12\bar\nabla_\nu h=0,
\]
where $\bar\nabla$ is the covariant derivative of the full NHEK metric $\bar g_{\mu\nu}$.
For an infinitesimal diffeomorphism,
\[
\delta_\xi\,\bar g_{\mu\nu}
=
\bar\nabla_\mu\xi_\nu+\bar\nabla_\nu\xi_\mu.
\]
The natural large rotational diffeomorphism has the schematic form
\[
\xi^{\mathrm{rot}}_n
=
H_n(x^i)\,\partial_\phi+\gamma_n(x)\bar \nabla^\mu H_n(x^i)\,\partial_\mu,
\]
where, $x^\mu$ denote coordinates on the full spacetime while $x^i$ are coordinates on the near-horizon $AdS_2$ part. The associated metric perturbation is
\[
h_{\mu\nu}^{\mathrm{rot}}
=
\mathcal L_{\xi_{\mathrm{rot}}}\bar g_{\mu\nu}.
\]
These perturbations are found to be traceless by construction. Demanding them to be physical, we impose the harmonic gauge and fix the unknown functions $\gamma_n(x)$ as follows,
\begin{align}
    \gamma_n(x)&=c_1+c_2\bigg(\cos\theta+2\log\left(\tan\frac\theta2\right)\bigg)+4\,\text{sgn}(n)\,\log\left(\sin\theta\right).
\end{align}
The compensating diffeomorphism denoted by
\begin{align}
    \hat\xi_n^{\rm rot}=\gamma_n(x)\bar\nabla^\mu H_n(x^a)\,\partial_\mu,
\end{align}
must be a small diffeomorphism and hence, normalizable. For normalizability one requires,
\begin{align}
    \begin{split}
        \langle\hat\xi_{-n}^{\rm rot}\,|\,\hat\xi_n^{\rm rot}\rangle&=\int d^4x\,\sqrt{\bar g}\,\,\,\hat\xi_{-n}^{\rm rot}{}^{\mu}\,\hat\xi_n^{\rm rot}{}_\mu<\infty.
    \end{split}
\end{align}
To meet the above criterion we rather demand the integrand to be regular over the entire range of integration. Despite choosing $c_2$ appropriately, the integrand fails to be regular at one of the poles. For the choice $c_2=-2\,\text{sgn}(n)$, the integrand is regular everywhere except at $\theta=\pi$. Similarly, when $c_2=2\,\text{sgn}(n)$, the integrand is found singular at $\theta=0$. It suggests one can never find a constant $c_2$ such that the integrand of the squared norm $\langle\hat\xi_{-n}^{\rm rot}\,|\,\hat\xi_n^{\rm rot}\rangle$ is regular over the interval $\theta\in[0,\pi]$.
Hence, the apparent compensating part of the gauge transformation, even though restores harmonic gauge, is non-normalizable. %can lead to singular behavior at the poles $\theta=0 $ or $\pi$. 
Thus, they can not be treated as small gauge transformation and will lead to the change in physical data of the fluctuation $h_{\mu \nu}.$ As it has been very nicely emphasized in~\cite{Kolanowski:2024zrq}, the above result simply implies that harmonic gauge is not a good gauge choice for this system\footnote{The harmonic gauge function transforms as
$
\delta_\xi \mathcal F_\mu
=
\left(\bar\nabla^2\delta_\mu{}^\nu+\bar R_\mu{}^\nu\right)\xi_\nu.$
Thus the Faddeev--Popov operator is
$
\mathcal M_\mu{}^\nu
=
\bar\nabla^2\delta_\mu{}^\nu+\bar R_\mu{}^\nu,$
which for the Ricci flat NHEK geometry reduces to
$
\mathcal M_\mu{}^\nu
=
\bar\nabla^2\delta_\mu{}^\nu.$ Since this operator has a non zero kernel, it is non invertible and hence it is not a good gauge choice.}. 

However, when studied from a dimensionally reduced 2-dimentional effective theory perspective, the corresponding lower dimensional rotational vector zero mode certainly exists. But its uplift are not naturally represented as smooth zero modes of the ordinary four-dimensional harmonic gauge operator. Thus the important conclusion is not that harmonic gauge is intrinsically invalid on Ricci-flat backgrounds. Rather, the problem is more specific: the combination of Ricci flatness, rotation, the near-horizon geometry, the large rotational diffeomorphism, and the desired regularity/boundary conditions produces a mismatch between the natural lower-dimensional rotational mode and the four-dimensional harmonic-gauge representative.

There are two possible ways to find a suitable gauge condition for the four dimensional system : 
\begin{itemize}
\item {\bf A bottom-up approach : }To make a four dimensional gauge choice by uplifting the two dimensional gauge choices. Although this leads to a non covariant gauge choice in the higher dimension (like Coulomb gauge in QED), one can try to find it it solves the system. 
\item {\bf A top-down approach : }Since we are studying the system around the extremal background, another possibility may be to write a gauge fixing condition involving the background metric and its symmetry data. This might lead to an invertible Faddeev-Popov operator. 
\end{itemize}
While both of these approaches look promising, a complete construction is still under construction.
We will report on this construction in a future study.

\section{The BMPV Black Hole}\label{section4}
Let us finally discus the 5-dimensional rotating BMPV black hole. This is the canonical supersymmetric rotating black-hole solution of five-dimensional asymptotically flat supergravity theory. Introduced in 1996, it provided one of the earliest examples in which a rotating black hole could be described both macroscopically in supergravity and microscopically in terms of D-brane states. The exact agreement between the Bekenstein--Hawking entropy and the microscopic degeneracy of BPS states made the BMPV solution an important milestone in the development of the string-theoretic interpretation of black-hole entropy.

\subsection{Different Constructions of the BMPV Solution}

An important feature of BMPV is that the same geometry can be obtained from several rather different viewpoints. This feature helps us in understanding the system well. Below we mention some of them, as that will be useful for the present paper.

\begin{itemize}
\item 
{\bf D-brane and string-theory construction :} The original BMPV solution was obtained in the context of string theory by starting with higher-dimensional configurations carrying momentum and brane charges and using boosts, compactification and string dualities. In the type-IIB description, it is naturally embedded in the D1--D5--P system: D1-branes wrap a common circle, D5-branes wrap the circle together with an internal four-manifold, and momentum propagates along the common direction. Rotation is then included while maintaining the BPS condition. This construction makes the microscopic interpretation of the solution particularly transparent \cite{Breckenridge:1996is}.

\item {\bf BPS limit of general charged rotating black holes :} BMPV can also be obtained by taking an appropriate supersymmetric limit of a more general non-extremal rotating charged black hole. Cveti\v{c} and Youm constructed broad families of rotating five-dimensional black holes by starting with a neutral rotating seed and applying solution-generating transformations associated with the duality group of the compactified theory \cite{Cvetic:1996xz}%CveticYoum}
. Their solutions possess independent mass, charges and two rotation parameters. Taking the BPS limit removes the non-extremality parameter and simultaneously constrains the angular momenta, yielding the BMPV branch.

Schematically,
\begin{equation}
\text{neutral Myers--Perry}
\;\longrightarrow\;
\text{charged Cveti\v{c}--Youm}
\;\longrightarrow\;
\text{BMPV}.
\end{equation}

The distinction between the extremal and BPS limits is important: a general charged rotating solution can possess both supersymmetric and non-supersymmetric extremal branches. If we seek for any generalization from the BMPV solution, this constraction makes it easier.

\item {\bf From CCLP to BMPV:} Finally, we present the construction that we have primarily used in the present work. BMPV can be embedded into the much more general family constructed by Chong, Cveti\v{c}, L\"u and Pope (CCLP) \cite{Chong:2005hr}%CCLP}
. 
The CCLP solution is the general non-extremal electrically charged rotating black hole of five-dimensional minimal gauged supergravity and contains four independent parameters: mass, charge and two angular momenta. The BMPV solution can be obtained in a particular limit from the generic CCLP solution. Starting from CCLP, one first takes the ungauged limit
\begin{equation}
g\rightarrow 0,
\end{equation}
where $g$ is the gauge coupling and $L_{\rm AdS_5}=g^{-1}$. This gives an asymptotically flat charged rotating solution of five-dimensional minimal supergravity. The 
Bosonic action of 5D minimal SUGRA in asymptotically flat space is given as,
\begin{equation}
S_{\rm EMCS}
=
\frac{1}{16\pi G_5}
\int d^5x
\left[
\sqrt{-g}
\left(
R-F_{\mu\nu}F^{\mu\nu}
\right)
-
\frac{2}{3\sqrt{3}}
\epsilon^{\mu\nu\rho\sigma\lambda}
A_\mu F_{\nu\rho}F_{\sigma\lambda}
\right].\label{5d_action}
\end{equation}

where, $g_{\mu\nu}$ is the 5D metric with signature $(-,+,+,+,+)$ and $R$ is the corresponding ricci scalar. $A_{\mu}$ is the gauge field and $F_{\mu\nu}$ is the associated field strength. The third term in the action is a topological Chern-Simons term, where, $\epsilon^{\mu\nu\rho\sigma\lambda}$ is the fully anti-symmetric Levi-civita symbol in 5D. It is a tensor density. The equations of motion for the metric and gauge fields are as follows,
\begin{align}\label{eom}
& EOM_{g}: \quad R_{\mu\nu}
- 2\left(
F_{\mu\rho}F_{\nu}{}^{\rho}
- \frac{1}{6} g_{\mu\nu}\,F^2
\right)
= 0, \\
& EOM_{A}: \quad \partial_\mu\left(\sqrt{-g}F^{\mu\nu}\right)
-\frac{1}{2\sqrt{3}}
\epsilon^{\nu\rho\sigma\gamma\delta}
F_{\rho\sigma}F_{\gamma\delta}
=0.
\end{align}

The solution of above equations gives an asymptotically flat charged rotating solution of five-dimensional minimal supergravity. One then imposes the BPS relation between the mass and charge parameters and restricts the two angular momenta to the supersymmetric BMPV combination. After appropriate parameter and coordinate redefinitions, the resulting metric and Maxwell field reduce to those of BMPV. Thus schematically,
\begin{equation}
\text{CCLP}
\;\xrightarrow{\,g\to0\,}\;
\text{asymptotically flat charged rotating black hole}
\;\xrightarrow{\rm BPS}\;
\text{BMPV}.
\end{equation}

This construction is useful because it places BMPV as a special corner of the full parameter space of charged, doubly rotating five-dimensional black holes. It is also useful for generalizing studies to the gauged case. 

\subsection*{Non-extremal Black Hole in minimal supergravity}

First, let us take the non-extremal solution in this category. We have to set $g=0$ (zero cosmological constant) and keep only one of the (left or right) angular momenta nonzero. The solution now has parameters: mass $m$, one angular momentum $a$, charge $Q$. The CCLP gauge field has to be shifted by a gauge transformation, so that it satisfies our EOMs. Thus, the CCLP non-extremal solution satisfying \eqref{eom} is given by: 
\begin{align}\label{CCLP}
& ds^2 = -\left(dt - \frac{2Q a\tilde\omega}{\rho^2}\right) dt - \frac{2Q a^2\tilde\omega^2}{\rho^2} + \frac{\tilde f}{\rho^4}(dt - a\tilde\omega)^2 +  \rho^2 \left(\frac{dr^2}{\Delta_r} + d\Omega_3^2 \right), \\
& A = \frac{\sqrt3}{2}dt - \frac{\sqrt3}{2}\frac{Q}{\rho^2} (dt - a\tilde\omega),
\end{align}
where,
\begin{align}\label{CCLP-funcs}
    &\tilde{\omega} = \sin^2\tilde\theta\,d\varphi-\cos^2\tilde\theta\,d\tilde\psi,\quad d\Omega_3^2=d\tilde\theta^2+\sin^2\tilde\theta\,d\varphi^2+\cos^2\tilde\theta\,d\tilde\psi^2,\\[5pt]
	& \rho^2 = r^2 + a^2, \quad \tilde f = 2m\rho^2 - Q^2,\quad  r^2\Delta_r = \rho^4 + Q^2 - 2a^2Q - 2m r^2.%\quad\textcolor{red}{(f\stackrel{?}{=}2m\rho^2-Q^2)}\\
\end{align}
Here, $t\in(-\infty,\infty)$, $r\in[0,\infty)$, $\tilde\theta\in[0,\frac\pi2]$, and $\varphi,\,\tilde\psi\in[0,2\pi)$. Coordinates $(\varphi,\tilde\psi)$ are periodic with a period of $2\pi$ each. Furthermore, the horizon and temperature are given by,
\begin{align}
	& 2mr_{+}^2 = (r_{+}^2 + a^2)^2 + Q^2 - 2a^2Q,\label{mass-r_+}\\
	& \kappa = 2\pi T = \frac{r_{+}^4 - (Q-a^2)^2}{r_{+}((r_{+}^2+a^2)^2 - a^2 Q%^2
    )}.\label{CCLP_Temp}
\end{align}
\subsection*{The BPS limit and BMPV solution}
In the BPS limit, mass is not an independent parameter but saturates the supersymmetric Bogomol'nyi bound,
\begin{equation}
M = Q.
\end{equation}

Imposing this condition in the above non-BPS solution, we then can get the BMPV solution by using the following radial transformation,
\begin{align}\label{CCLP-BMPV-coord}
r^2 = R^2 + Q - a^2, \quad H = 1 + \frac{Q}{R^2}.
\end{align}
The horizon in CCLP coordinate is given as,
\begin{align}
r_0^2 = Q - a^2 \implies R = 0, \quad \kappa = 2 \pi T= 0.
\end{align}
Thus the full BMPV solution metric and gauge field solutions are given as,
\begin{align}\label{BMPV}
& ds^2
=
-H^{-2}(dt+\hat\omega)^2
+
H (dR^2 + R^2 d\Omega_3^2), \\
& A
=
\frac{\sqrt{3}}{2}
H^{-1}(dt+\hat\omega),
\end{align}
where
\begin{equation}
H(r)=1+\frac{Q}{R^2}, \quad \hat\omega = \frac{J}{2R^2}\,\tilde{\omega}, \quad \tilde{\omega} = \sin^2\tilde\theta\,d\varphi
-
\cos^2\tilde\theta\,d\tilde\psi, \quad J= 2 a Q.
\end{equation}
The horizon is located at $R = 0$. The horizon area is proportional to
\begin{equation}
A_H\propto\sqrt{4Q^3-J^2}.
\end{equation}
Further the regularity of the horizon requires
\begin{equation}
J^2 \leq 4Q^3, \quad \text{or equivalently} \quad \ a^2 \leq Q.
\end{equation}

A remarkable property of BMPV is that it carries finite angular momentum while remaining BPS. In five dimensions, the three sphere has $SO(4)\simeq SU(2)_L\times SU(2)_R$ isometry and a generic rotating black hole admits two independent angular momenta $J_1$ and $J_2$. For BMPV these are constrained to have equal magnitude, conventionally
\begin{equation}
J_1=J_2=J.
\end{equation}
The equality enhances the rotational symmetry to $SU(2)\times U(1)$.  Consequently the BMPV solution is extremal, although extremality by itself does not imply supersymmetry.
We end this section with a comparative discussion on the BMPV solution that we work with in this work with the D1-D5-p system and the generic Cveti\v{c}--Youm  solution. The D-brane system is beneficial for the microscopic interpretations and in general contains three different charges and an angular momentum. Similarly, 
  the Cveti\v{c}--Youm solution is a more general non-extremal,
    three-charge, two-spin rotating black hole, whose BPS limit gives 
    the three-charge BMPV black hole. The minimal BPMV solutions that we work with is a special case of the both systems, where all the three charges are set to the same value. It would be nice to study the generic BMPV black holes, where the charges are kept unconstrained. 
\end{itemize}
\subsection{Near-Extremal extension of BMPV Black Holes}\label{sec:log T near-BMPV}
In this section, we systematically present the leading quantum logarithmic corrections in the temperature to the entropy for the near-extremally extended minimal BMPV black holes. We closely follow the strategy developed in \cite{Banerjee:2023quv}, suitably extending it to the present system. In \cite{Sen:2008vm,Sen:2012kpz}, Sen has shown that for asymptotically flat extremal geometries, the entropy can be obtained from the near horizon geometry of the extremal black holes. In particular, to find the leading quantum logarithmic corrections to the entropy, the computation boils down to the evaluation of the contributions of the quadratic fluctuations round the near horizon extremal geometry to the partition function. For finding the $\log T $ correction to the extremal entropy one needs to fist find the near-extremal extension of the extremal near horizon geometry. As was explained in \cite{Banerjee:2023quv,Banerjee:2023gll}, the required contribution then comes from only the extremal zero modes in the corresponding near horizon background. Here we first present a systematic derivation of the near horizon geometry of the BMPV solution and its near-extremal extension. Next we construct the corresponding extremal zero modes.

\subsection*{Near-horizon BMPV solution and its near-extremal extension}
Imposing the limit $R\to 0$ on the full BMPV solution \eqref{BMPV}, succeeded by the following re-definitions,
\begin{align}
    &\rho = R^2,\quad t = \frac{Q^{3/2}}{2}\tau, \quad Q \equiv 4q^2,
\end{align}
we end up with the near-horizon BMPV solution,
\begin{align}
    & ds^2 = -\left(q\rho d\tau + \frac{J}{8q^2}\tilde{\omega}\right)^2 + q^2 \frac{d\rho^2}{\rho^2} + 4q^2 d\Omega_3^2, \label{NH_BMPV_Hopf}\\[5pt]
    & A = \frac{\sqrt{3}}{2}\left(q\rho d\tau + \frac{J}{8q^2}\tilde{\omega}\right),
\end{align}
where,
\begin{align}
    \ell_{AdS_2} = q\,,\qquad&\text{and}\qquad \tilde{\omega} = \sin^2{\tilde\theta}\,d\varphi - \cos^2\tilde\theta\,d\psi.
\end{align}
This too solves the EOMs \eqref{eom}. In $AdS_2$--adapted coordinates, the same near-horizon geometry in Euclidean signature takes the following form,
\begin{align}\label{BMPV_Kerr-CFT}
\begin{split}
    &ds^2=\frac{Q}{4}\left[\sinh^2\eta\,d\theta^2+d\eta^2+d\hat\theta^2+\sin^2\hat\theta\, d\phi^2+\cos^2\omega\left(d\psi+\cos\hat\theta\,d\phi+i\tan\omega(\cosh\eta-1)d\theta\right)^2\right],\\[5pt]
    &A_\mu\,dx^\mu
    =-\frac{\sqrt{3Q}}{4}\,\bigg[i\,\cos\omega\big(\tan^2\omega+\cosh \eta\big)\,d\theta-\sin\omega\big(\cos \hat\theta\,d\phi+d\psi\big)\bigg],
    \end{split}
\end{align}
where, $a=-\sqrt{Q}\sin\omega$ and $0\leq\theta<2\pi$, $0\leq\eta<\infty$, $0\leq\hat\theta\leq\pi$, $0\leq\phi<2\pi$, and $0\leq\psi<4\pi$. Coordinates $(\theta,\phi,\psi)$ are periodic with the following periodicities $\theta\sim\theta+2\pi$, $\phi\sim\phi+2\pi$, and $\psi\sim\psi+4\pi$. The map from \eqref{NH_BMPV_Hopf} to \eqref{BMPV_Kerr-CFT} involves a sequence of non-trivial coordinate transformations and has been discussed in details in appendix \ref{app:Kerr CFT form}. The near horizon geometry has an enhancement of symmetry to $SL(2,\mathbb{R})\times SU(2)\times U(1) \times U(1)$, as compared to the full geometry. Here, The $SL(2,\mathbb{R})$ corresponds to isometries of $AdS_2$, $SU(2)$ are the isometries of $S^2$, $U(1)$ is the isometry along the rotational axis and the last $U(1)$ corresponds to the electric charge of the gauge field. The $SL(2,\mathbb{R})$ isometry is generated by the following set of Killing vector\footnote{Once rotation is turned off or $a=0$, these boil down to the global isometry generators of $AdS_2$.} fields,
\begin{align}
    \xi_0&=i\,\partial_\theta+\frac{a}{\sqrt{Q-a^2}}\,\partial_\psi,\quad\xi_{\pm}=e^{\pm i \theta}\left[\pm i\coth\eta\,\partial_\theta+\partial_\eta\pm\frac a{\sqrt{Q-a^2}}\tanh\left(\frac\eta2\right)\partial_\psi\right],\label{eq:generators AdS2 isometry NH-BMPV}
\end{align}
satisfying the $\mathfrak{sl}(2,\mathbb R)$ algebra,
\begin{align}
    [\xi_0,\xi_{\pm}]&=\mp\,\xi_\pm,\quad[\xi_+,\xi_-]=-2\,\xi_0.
\end{align}
Since, at the Lie algebra level $\mathfrak{sl}(2,\mathbb R)\simeq \mathfrak{so}(2,1)$, the above generators maybe equivalently represented in the $\mathfrak{so}(2,1)$ basis as,
\begin{align}
    \tilde\xi_0&=\xi_0,\quad\tilde\xi_1=\frac{\xi_++\xi_-}2,\quad\tilde\xi_2=\frac{\xi_+-\xi_-}2,
\end{align}
which satisfy the algebra,
\begin{align}
    [\tilde\xi_0,\tilde\xi_1]=-\,\tilde\xi_2,\quad[\tilde\xi_1,\tilde\xi_2]=\tilde\xi_0,\quad[\tilde\xi_2,\tilde\xi_0]=\tilde\xi_1.
\end{align}
The $SO(4)\simeq SU(2)_L\times SU(2)_R$ symmetry group associated with the 3-sphere $S^3$ in 5D is spontaneously broken into $SU(2)_L\times U(1)$ by the near-horizon, BMPV saddle. The generators $\{L_I\}$ of the chiral $SU(2)_L$ sector and their associated algebra are as follows,
\begin{align}\label{SU2generator}
    \begin{split}
        &L_0=\partial_\phi\,,\quad L_1=-\sin{\phi}\,\partial_{\hat\theta}-\cot\hat\theta\,\cos\phi\,\partial_\phi+\csc\hat\theta\,\cos\phi\,\partial_{\psi}\,,\\[5pt]
        &L_2=\cos{\phi}\,\partial_{\hat\theta}-\cot\hat\theta\,\sin\phi\,\partial_\phi+\csc\hat\theta\,\sin\phi\,\partial_{\psi},\\[5pt]
        &[L_I,L_J]=\epsilon_{IJK}L_K,\qquad\epsilon_{210}=1,\qquad I,J,K\in\{0,1,2\}.
    \end{split}
\end{align}
Here, $\epsilon_{IJK}$ is the fully anti-symmetric Levi-Civita symbol. Similarly, the generators $\{R_I\}$ of the chiral $SU(2)_R$ sector along with their algebra are given by,
\begin{align}
    \begin{split}
        &R_0=\partial_\psi,\quad R_1=-\sin\psi\,\partial_{\hat\theta}+\csc\hat\theta\,\cos\psi\,\partial_{\phi}-\cot\hat\theta\,\cos\psi\,\partial_\psi\label{eq:rot iso generator}\\[5pt]
        & R_2=\cos\psi\,\partial_{\hat\theta}+\csc\hat\theta\,\sin\psi\,\partial_\phi-\cot\hat\theta\,\sin\psi\,\partial_\psi,\\[5pt]
        &[R_I,R_J]=\epsilon_{IJK}R_K,\qquad \epsilon_{210}=1,\qquad I,J,K\in\{0,1,2\}.
    \end{split}
\end{align}
Introducing a non-trivial rotation or equivalently the choice of saddle, spontaneously breaks the $SU(2)_R$ symmetry group into a $U(1)$. It is easy to check\footnote{It can be shown that although $\mathcal L_{\tilde\xi_{1,2}}\bar A_\mu\neq0$, $\mathcal L_{\tilde\xi_{1,2}}\bar F_{\mu\nu}=0$. Hence, one can always find a suitable gauge parameter that compensates the non-trivial Lie-dragging of the near-horizon, extremal background $U(1)$ gauge field. This has been discussed in details in appendix \ref{Lie_drag_U(1)}. In this sense, $\mathcal L_V\bar \Psi=0$.} that,
\begin{align}
        \mathcal{L}_{V} \bar\Psi=0,\quad V\in\{\{\tilde\xi_i\},\{L_I\},R_0\},\quad i,I\in\{0,1,2\},\qquad\mathcal L_{R_{1,2}} \bar\Psi\neq0,
\end{align}
where, $\bar\Psi$ denotes a generic field in the near-horizon BMPV background. Thus, the $U(1)$ in $SU(2)_L\times U(1)$ corresponds to the rotational isometry generated by the vector field $\partial_\phi$\footnote{It is the vector dual to the invariant 1-form $\hat\sigma_3=d\psi+\cos\hat\theta\,d\phi$ in \eqref{BMPV_Kerr-CFT}.}.
\subsubsection*{Near extremal extension of BMPV}
%The near-extremal mass and horizon radius maybe expanded perturbatively in $T$ (while J is fixed at its extremal value) as shown below,
%\begin{align}
    %\begin{split}
        %r_+(T)&=\sqrt {Q - 
        %a^2}  +  \frac {\pi Q T} {2}+\frac {7\pi^2 Q^2 T^2} {8\sqrt {Q - a^2}}\\[5pt]
        %m(T)&=Q+ \frac {1} {2}\pi^2 Q^2 T^2 + \frac {3\pi^3 Q^3 T^3} {2\sqrt {Q - 
      %a^2}} 
    %\end{split}
%\end{align}
Just above extremality, the horizon radius $r_+$ is close to $r_0=\sqrt{Q-a^2}$, the extremal horizon radius. For a small non-zero temperature $T$, we may expand the horizon radius $r_+$ about $r_0$ in small $T$ as follows,
\begin{align}
    \begin{split}
        r_+(T)&=r_0+A\,T+B\,T^2,
    \end{split}
\end{align}
where, the charge $Q$ (or, angular momentum $J$) is held fixed at its extremal value. Replacing $r_+$ with this ansatz back in \eqref{CCLP_Temp}, we expand the RHS in small $T$ and compare coefficients from either sides to obtain,
\begin{align}
    \begin{split}
        A=\frac{\pi\,Q}{2}\,,\qquad\text{and}\qquad B=\frac{7\pi^2\,Q^2}{8\,r_0}.
    \end{split}
\end{align}
Using \eqref{mass-r_+}, the mass parameter $m$ can be similarly expanded in small $T$, however, correct only up to $O(T^2)$. In terms of the dimensionless perturbative parameter $\epsilon(T)$, the near-extremal mass and horizon radius are as follows,
\begin{align}
    \begin{split}\label{pert_exp}
        r_+(\epsilon)&=r_0\Big[1  +  \frac {\pi} {2}\,\epsilon+\frac {7\,\pi^2}{8}\,\epsilon^2\Big]+\mathcal{O}(\epsilon^3)\,,\quad \epsilon(T)=\dfrac{Q\,T}{\sqrt{Q-a^2}}\ll1\\[5pt]
        m(\epsilon)&=Q\Big[1+ \frac {\pi^2} {2}\bigg(1-\frac{a^2}{Q}\bigg) \epsilon^2\,\Big] + \mathcal{O}(\epsilon^3).
    \end{split}
\end{align}
To zoom into the near-horizon throat of the near-BMPV solution, we replace the black hole parameters with their respective near-extremal expansions and subject the non-extremal configuration of fields in \eqref{CCLP} to the following change of coordinates\footnote{The change of coordinates aligns with that used in \cite{Bardeen:1999px}.},
\begin{align}
    \begin{split}
        t(\theta)&=\frac{Q\,\theta}{2\pi r_0\epsilon}\,,\quad r(\eta)=r_+(\epsilon)+\frac{\pi }{2}r_0\epsilon\left(\cosh\eta-1\right)\,,\quad\chi(\theta,\psi)=\psi-\frac{a\,\theta}{\sqrt{Q-a^2}}.\label{NH_coord_transf}
    \end{split}
\end{align}
%Perturbatively, 
The transformed fields may be expanded perturbatively up to $\mathcal{O}(\epsilon)$ to obtain the near-extremal configuration. Thus, the near-extremal metric is given by, $g_{\mu\nu}=g_{\mu\nu}^{(0)}+\,\epsilon g_{\mu\nu}^{(1)}$. In Euclidean signature $(\theta\rightarrow-i\theta)$, the $\mathcal{O}(1)$ piece gives back the BMPV solution in \eqref{BMPV_Kerr-CFT}, 
\begin{align}
    \begin{split}
        ds^2&=\frac{Q}{4}\bigg[\sinh^2\eta \,d\theta^2+d\eta^2+d\hat\theta^2+\sin^2\hat\theta\, d\phi^2+\cos^2\omega\,\big(d\psi+\cos\hat\theta\,d\phi+i\tan\omega\,(\cosh\eta-1)d\theta\big)^2\bigg]
    \end{split}
\end{align}
while the components $g^{(1)}_{\mu\nu}$ in the $\mathcal{O}(\epsilon)$ correction are given by,
\begin{align}
    \begin{split}
        g_{\theta\theta}^{(1)}&=-\frac {\pi} {8}Q\,(\cosh \eta - 
      1)\left (\cos^2\omega (2\cos2\omega + 1)\cosh^2\eta + \left (3 - 
         4\sin^4\omega \right)\cosh \eta + 11\sin^2\omega - 
     6 \right)\\[10pt]
     g_{\theta\phi}^{(1)}&=i\frac{\pi} {32}\,Q\cos \hat\theta\,\sin 2\omega\, (\cosh \eta - 
    1)\left (-11+\left (-7+8\sin^2\omega\right)\cosh \eta  \right)=g_{\phi\theta}^{(1)}\\[10pt]
    g_{\theta\psi}^{(1)}&=i\frac{\pi} {32}\,Q\sin 2\omega\, (\cosh \eta - 
    1)\left (-11+\left (-7+8\sin^2\omega\right)\cosh \eta  \right)=g_{\psi\theta}^{(1)}\\[10pt]
    g_{\eta\eta}^{(1)}&=\frac{\pi \,Q\left (-6+(2+\cos 2\omega)(\cosh\eta+1)\cosh\eta\right)} {8\,(\cosh \eta + 1)}\\[10pt]
%      \end{split}
% \end{align*}
% \begin{align}
%     \begin{split}
g_{\hat\theta\hat\theta}^{(1)}&=\frac {\pi} {4}\,Q\cos^2\omega\,\cosh \eta\,,\quad g_{\phi\phi}^{(1)}=\frac {\pi} {4}\, 
Q\cos^2\omega\,\left (1 + 
   2\sin^2\omega\, \cos^2\hat\theta  \right)\cosh \eta\\[10pt]
   g_{\phi\psi}^{(1)}&=\frac {\pi} {4}\, 
Q\cos^2\omega\left (1 + 2 \sin^2\omega \right)\cos \hat\theta\, \cosh \eta=g_{\psi\phi}^{(1)}\,,\quad g_{\psi\psi}^{(1)}=\frac {\pi} {4}\, 
Q\cos^2\omega\left (1 + 2 \sin^2\omega \right)\cosh \eta
    \end{split}
\end{align}
where, $a=-\sqrt{Q}\sin\omega$. It is important to note that the near-horizon, near-BMPV metric is both regular and non-singular (i.e, det $g_{\mu\nu}\neq0$) in the limits $\eta\rightarrow0$ and $\epsilon\rightarrow 0$. Similarly, the $\mathcal{O}(1)$ and $\mathcal{O}(\epsilon)$ pieces of the near-extremal gauge field $A_\mu=A^{(0)}_\mu+\epsilon A^{(1)}_\mu$ are,
\begin{align}
    \begin{split}
        A_\mu^{(0)}dx^\mu
        &=-\frac{\sqrt{3Q}}{4}\,\bigg[i\,\cos\omega\big(\tan^2\omega+\cosh \eta\big)\,d\theta-\sin\omega\big(\cos \hat\theta\,d\phi+d\psi\big)\bigg]
    \end{split}
\end{align}
and,
\begin{align}
    \begin{split}
        \epsilon A_\mu^{(1)}dx^\mu&=-\frac{\pi\sqrt{3Q}}{4}\,\epsilon\bigg[\,\frac i 4\,\big(7-4 \sin^2\omega\cosh \eta - \left(3-4\sin^2\omega\right)\cosh^2 \eta\big)\,d\theta\\[5pt]
        &\qquad\qquad\qquad\qquad+\,\frac{\sin 2\omega}{2}\big (d\psi+\cos \hat\theta\, d\phi\big)\cosh \eta\bigg]\times\cos\omega
    \end{split}
\end{align}
The isometry group of the near-horizon, BMPV background is partially preserved when a small, non-zero temperature is turned on. The $AdS_2$ isometry group $SL(2,\mathbb{R})$ reduces to a $U(1)$\footnote{In the near-horizon, near-BMPV background, $\mathcal L_{\tilde \xi_0}\bar \Psi=0$ while $\mathcal L_{\tilde \xi_{1,2}}\bar \Psi\neq0$, where, $\bar\Psi$ represents a generic field in the chosen background.} in the near-horizon, near-BMPV background. Hence, the isometry group of the near-horizon, near-BMPV background is $U(1)\times SU(2)\times U(1)\times U(1)$. 

\subsection{Construction of extremal zero modes}
Here we lay out the strategy to find the zero modes of the system in the near-horizon extremal background. The prescription is covariant and hence can be suitably adopted for any system. To find the zero modes, we first expand around our background solution
\begin{equation}
g_{\mu\nu}
=
\bar g_{\mu\nu}+h_{\mu\nu},
\qquad
A_\mu
=
\bar A_\mu+a_\mu.
\end{equation}

Here $\bar g_{\mu\nu}$ and $\bar A_\mu$ are the BMPV background presented in \eqref{BMPV_Kerr-CFT} and $h_{\mu \nu}$ and $a_\mu$ are some fluctuations around it. All barred quantities are evaluated with respect to the background. Since the background satisfies the full Einstein--Maxwell--Chern--Simons
equations of motion, we have 
\begin{equation}
\left.
EOM_g=\frac{\delta S_{\rm EMCS}}{\delta g_{\mu\nu}}
\right|_{\bar g,\bar A}
=0,
\qquad
EOM_A=\left.
\frac{\delta S_{\rm EMCS}}{\delta A_\mu}
\right|_{\bar g,\bar A}
=0.
\label{eq:backgroundEOM}
\end{equation}

\subsubsection*{Gauge fixing the fluctuations}

The Einstein-Maxwell-Chern-Simons theory have diffeomorphism and U(1) gauge invariance as its gauge symmetry. For the computation of the partition function, we first need to gauge fix the theory. For the gravitational perturbation we choose harmonic, or de Donder, gauge,
\begin{equation}
\mathcal F_\mu[h]
=
 \nabla^\nu h_{\mu\nu}
-\frac12 \nabla_\mu h,
\qquad
h=g^{\mu\nu}h_{\mu\nu}.
\label{eq:harmonicGF}
\end{equation}

The corresponding gauge-fixing action is
\begin{equation}
S_{\rm gf}^{(g)}
=
-\frac{1}{32\pi G_5}
\int d^5x\,\sqrt{- g}\,
\mathcal F_\mu[h]\mathcal F^\mu[h].
\label{eq:metricGFaction}
\end{equation}

For the gauge-field perturbation we impose Lorenz gauge,
\begin{equation}
\mathcal G[a]
=
 \nabla^\mu a_\mu,
\label{eq:LorenzGF}
\end{equation}
with
\begin{equation}
S_{\rm gf}^{(A)}
=
-\frac{1}{8\pi G_5}
\int d^5x\,\sqrt{- g}\,
\mathcal G[a]^2.
\label{eq:gaugeGFaction}
\end{equation}

Thus the quadratic gauge-fixed theory is built from
\begin{equation}
S_{\rm tot}^{(2)}
=
S_{\rm EMCS}^{(2)}
+
S_{\rm gf}^{(g,2)}
+
S_{\rm gf}^{(A,2)},
\end{equation}
keeping only quadratic terms in the fluctuations. Thus every other terms need to be evaluated on the background $(\bar g, \bar A).$

Next we consider Pure-gauge fluctuations of this system via an infinitesimal diffeomorphism generated by $\xi^\mu$ together
with an Abelian gauge transformation generated by $\Lambda$. Keeping the quadratic action in mind. we have the metric fluctuation is given by,
\begin{equation}
h_{\mu\nu}^{(\xi)}
=
\mathcal L_\xi \bar g_{\mu\nu}
=
\bar \nabla_\mu\xi_\nu+\bar \nabla_\nu\xi_\mu,
\label{eq:puregaugemetric}
\end{equation}
where $\mathcal L_\xi$ denotes Lie-drag of any field along a vector $\xi$. 
Similarly, the gauge-field fluctuation is
\begin{equation}
a_\mu^{(\xi,\Lambda)}
=
\mathcal L_\xi\bar A_\mu+\bar \nabla_\mu\Lambda.
\label{eq:puregaugeA}
\end{equation}

The Lie derivative of the background gauge field can be written as
\begin{align}
(\mathcal L_\xi\bar A)_\mu
&=
\xi^\nu\bar \nabla_\nu\bar A_\mu
+
\bar A_\nu\bar \nabla_\mu\xi^\nu
\nonumber\\
&=
\xi^\nu\bar F_{\nu\mu}
+
\bar \nabla_\mu(\xi^\nu\bar A_\nu).
\label{eq:CartanComponents}
\end{align}

Therefore,
\begin{equation}
a_\mu^{(\xi,\Lambda)}
=
\xi^\nu\bar F_{\nu\mu}
+
\bar \nabla_\mu
\left(
\xi^\nu\bar A_\nu+\Lambda
\right).
\label{eq:aCovariant}
\end{equation}

The function $\Lambda$ is so far the independent $U(1)$ gauge parameter.

\subsubsection*{Definition of the zero modes and Residual gauge conditions}
Let us now define the zero modes. Up to boundary terms that are appropriately taken care by the GHY terms, the gauge-invariant quadratic action is schematically\footnote{Look at appendix \ref{APPC} for detailed derivation.}
\begin{equation}
S_{\rm tot}^{(2)}
=-\frac{1}{32\pi G_5}\int d^5x\,\sqrt{\bar g}
\left[
 h_{\mu\nu}\,\delta E^{\mu\nu}[h,a]
+a_\mu\,\delta E_A^\mu[h,a]
\right].
\end{equation}
The linearized equations split as
\[
\delta E^{\mu\nu}[h,a]
=
\delta_g E^{\mu\nu}[h]
+
\delta_A E^{\mu\nu}[a], \quad
\delta E_A^\mu[h,a]
=
\delta_g E_A^\mu[h]
+
\delta_A E_A^\mu[a].
\]
Therefore
\begin{align}
S^{(2)}_{\rm tot}
=
-\frac12\int d^5x\,\sqrt{\bar g}\Big[
&h_{\mu\nu}\,\delta_g E^{\mu\nu}[h]
+h_{\mu\nu}\,\delta_A E^{\mu\nu}[a]
+
a_\mu\,\delta_g E_A^\mu[h]
+a_\mu\,\delta_A E_A^\mu[a]
\Big].
\end{align}
Let us define
$ \,
S^{(2)}_{tot}=S^{(2)}_{hh}+S^{(2)}_{ha}+S^{(2)}_{aa} \,,
$
with
\[
S_{hh}^{(2)}
=
-\frac12
\int d^5x\,\sqrt{\bar g}\,
h_{\mu\nu}\,\delta_g E^{\mu\nu}[h],
\quad
S_{aa}^{(2)}
=
\frac12
\int d^5x\,\sqrt{-\bar g}\,
a_\mu\,\delta_A E_A^\mu[a],
\]
and
\[
S_{ha}^{(2)}
=
-\frac12
\int d^5x\,\sqrt{\bar g}
\left[
h_{\mu\nu}\,\delta_A E^{\mu\nu}[a]
+
a_\mu\,\delta_g E_A^\mu[h]
\right].
\]

Because the two mixed terms come from the same second variation of the action,
the corresponding operators are adjoints of one another, up to boundary terms:
\[
\int \sqrt{\bar g}\,
h_{\mu\nu}\,\delta_A E^{\mu\nu}[a]
=
\int \sqrt{\bar g}\,
a_\mu\,\delta_g E_A^\mu[h]
+\text{boundary}.
\]

If the boundary contribution vanishes or is treated appropriately, then
\[
S_{ha}^{(2)}
=-
\int \sqrt{\bar g}\,
h_{\mu\nu}\,\delta_A E^{\mu\nu}[a]
=
-\int \sqrt{\bar g}\,
a_\mu\,\delta_g E_A^\mu[h].
\]

Introducing the Block-Hessian form, we further write
\[
S^{(2)}
=
\frac12
\begin{pmatrix}
h & a
\end{pmatrix}
\begin{pmatrix}
\mathcal H_{gg} & \mathcal H_{gA}\\
\mathcal H_{Ag} & \mathcal H_{AA}
\end{pmatrix}
\begin{pmatrix}
h\\
a
\end{pmatrix}.
\]

The blocks are defined by
\[
\mathcal H_{gg}h=\delta_g E[h],
\qquad
\mathcal H_{gA}a=\delta_A E[a],
\]
\[
\mathcal H_{Ag}h=\delta_g E_A[h],
\qquad
\mathcal H_{AA}a=\delta_A E_A[a].
\]
The fluctuation modes are defined as zero modes, if they satisfy,
\begin{equation}
\delta E^{\mu\nu}[h,a]
=
\mathcal H_{gg}^{\mu\nu\,\rho\sigma}h_{\rho\sigma}
+
\mathcal H_{gA}^{\mu\nu\,\rho}a_\rho=0, \qquad
\delta E_A^\mu[h,a]
=
\mathcal H_{Ag}^{\mu\,\rho\sigma}h_{\rho\sigma}
+
\mathcal H_{AA}^{\mu\rho}a_\rho.
\end{equation}
As we have explicitly shown in the appendix, if the fluctuations satisfy the residual gauge conditions, they are the zero modes of the quadratic kinetic Hessian operator.
Hence we next implement the residual gauge conditions. For the metric fluctuations 
\begin{equation}
h_{\mu\nu}
=
\bar \nabla_\mu\xi_\nu+\bar \nabla_\nu\xi_\mu,
\end{equation}
the trace is
\begin{equation}
h
=
2\bar \nabla_\rho\xi^\rho.
\end{equation}

Using the definition
\begin{equation}
\mathcal F_\mu[h]
=
\bar \nabla^\nu h_{\mu\nu}
-\frac12\bar \nabla_\mu h,
\end{equation}
we obtain
\begin{align}
\mathcal F_\mu[h^{(\xi)}]
&=
\bar \nabla^\nu
\left(
\bar \nabla_\mu\xi_\nu
+
\bar \nabla_\nu\xi_\mu
\right)
-
\bar \nabla_\mu\bar \nabla_\rho\xi^\rho
\nonumber\\
&=
\bar \nabla^\nu\bar \nabla_\nu\xi_\mu
+
\bar \nabla^\nu\bar \nabla_\mu\xi_\nu
-
\bar \nabla_\mu\bar \nabla^\nu\xi_\nu.
\end{align}

Using
\begin{equation}
\left[
\bar \nabla^\nu,\bar \nabla_\mu
\right]\xi_\nu
=
\bar R_{\mu\nu}\xi^\nu,
\end{equation}
we find
\begin{equation}
\mathcal F_\mu[h^{(\xi)}]
=
\bar \square\xi_\mu
+
\bar R_{\mu\nu}\xi^\nu.
\label{eq:Pxi}
\end{equation}

Preservation of harmonic gauge therefore requires
\begin{equation}
P_\mu[\xi] \equiv \bar \square\xi_\mu
+
\bar R_{\mu\nu}\xi^\nu=0.
\end{equation}

Similarly imposing the residual Lorenz gauge condition for the gauge-field fluctuation,
\begin{equation}
a_\mu
=
\mathcal L_\xi\bar A_\mu+\bar \nabla_\mu\Lambda,
\end{equation}
the Lorenz-gauge functional becomes
\begin{align}
\mathcal G[a]
&=
\bar \nabla^\mu a_\mu
\nonumber\\
&=
\bar \nabla^\mu
(\mathcal L_\xi\bar A_\mu)
+
\bar \square\Lambda.
\end{align}
The residual Lorenz-gauge condition becomes
\begin{equation}
Q[\xi,\Lambda]
\equiv
\bar \square\lambda
+
\bar \nabla^\mu
(\mathcal L_\xi\bar A_\mu)=0,
\end{equation}
or equivalently the $\Lambda$ parameter gets fixed via following condition
\begin{equation}
\bar \square\Lambda
=
-
\bar\nabla^\mu
(\mathcal L_\xi\bar A_\mu).
\label{eq:residualLorenz}
\end{equation}

Using Eq.~\eqref{eq:CartanComponents}, this can also be written as
\begin{align}
Q[\xi,\Lambda]
&=
\bar \nabla^\mu
\left(
\xi^\nu\bar F_{\nu\mu}
\right)
+
\bar \square
\left(
\Lambda+\xi^\nu\bar A_\nu
\right).
\label{eq:Qalternative}
\end{align}

This equation shows explicitly how the $U(1)$ gauge parameter $\Lambda$
is required in order for the Lie-dragged gauge-field fluctuation to remain
in Lorenz gauge. An interesting scenario occurs when $\mathcal L_\xi\bar F_{\mu \nu}$ becomes zero for a diffeomorphism $\xi$. In such cases one can always choose a suitable U(1) gauge transformation parameter $\Lambda$ so that $a_\mu
=
\mathcal L_\xi\bar A_\mu+\bar \nabla_\mu\Lambda=0$\footnote{see appendix \ref{Lie_drag_U(1)} for details.}. In such cases, the gauge fluctuations gets completely decoupled from the system. In our constructions of the zero modes below, the above covariant discussions play a major role. 

\subsubsection{Different kinds of zero modes}
The zero modes are originated from enhanced isometries of the Euclidean BMPV solution. As we have already identified earlier, the global isometries of the near-horizon BMPV solutions are $SL(2,\mathbb{R})\times SU(2)\times U(1) \times U(1)$. All of these symmetries get enhanced near the $AdS_2$ boundary as they can be interpreted as large diffeomorphisms of $AdS_2$ metric, and large gauge transformations corresponding to the gauge groups. All of these zero modes are orthogonal to each other and hence do not mix among themselves. In particular, both zero and non-zero eigenmodes of the extremal kinetic operator have been normalized using the following orthonormality relation
\begin{align}
    \begin{split}
        \langle f_m|f_n\rangle&=\int d^5x\,\sqrt{\bar g}\,\,G_{IJ}\,f^{I}_m(x)f^J_n(x)=\delta_{m+n,0}\,,
    \end{split}
\end{align}
where, $G_{IJ}$ is the metric defined on the space of eigen functions induced by the extremal background spacetime metric $\bar g_{\mu\nu}$, and $f^I_m(x)$ denotes all kinds of basis eigenfunctions. Here, indices $(I,J)$ capture the tensorial structure of the fields spanned in the eigenbasis $\{f^I_m\}$. The forms of the zero modes are written by considering appropriate discrete basis elements in $AdS_2$ and $S^3$ and then suitably uplifting them to 5D. To be precise, the fluctuation fields are expressed in terms of the eigen basis functions as, $$h_{\mu\nu}= \sum_n c_n\, h^{(n)}_{\mu\nu}, \quad a_\mu= \sum_n b_n\,a^{(n)}_\mu.$$ 
While the forms of the basis functions $\big(h^{(n)}_{\mu\nu},a^{(n)}_\mu\big)$ are explicitly fixed, as we present below, the modes $(c_n, b_n)$ are path integrated over a suitably chosen complex steepest-descent contour. Next we present the explicit form of the BMPV zero modes.\\[200pt]
{\bf Schwarzian modes}\\[10pt]
The non-normalizable vector field generating a Schwarzian mode is given by,
\begin{align}
\begin{split}
    &\xi^{\rm Sch.}_n=\frac{e^{in\theta}\tanh^{| n |}\left (\frac {\eta} {2}\right)}{\sqrt{|n|(n^2-1)}}\times\bigg[i \big(n\csch^2\eta\,(| n | +\cosh \eta) + \text{sgn}(n)\big)\,\partial_\theta+| n |  \csch\eta\, (| n | +\cosh\eta)\,\partial_\eta\\[5pt]
    &\qquad\qquad\qquad\qquad\qquad\qquad+\frac {a\,\text{sgn} (n)\big (1+| n |  - n^2+\cosh \eta\big)} {\sqrt {Q -a^2}\,(1 + \cosh \eta)}\,\partial_{\psi}\Bigg],\qquad n\in\mathbb{Z},\,|n|\geq2.%\\[10pt]
    %&\text{where,}\qquad c_1=\frac{1}{2\,\pi^{3/2}\,Q(Q-a^2)^{1/4}}
\end{split}
\end{align}
For $n=0,\pm1$, the above expression reproduces the Killing vectors in \eqref{eq:generators AdS2 isometry NH-BMPV} upto an overall constant. To maintain the symmetry of the BMPV background, the vector field now depends on the rotation parameter $a$. Although the vector field $\xi^{\rm Sch.}_n$ Lie-drags both the extremal background metric and gauge fields non-trivially, the Lie dragging of the extremal background field strength $\bar F^{(0)}_{\mu\nu}$ along $\xi^{\rm\,Sch.}_n$ vanishes,
\begin{align}
    \begin{split}
        \mathcal{L}_{\xi^{\rm Sch.}_n}\,\bar{F}^{(0)}_{\mu\nu}&=0,
    \end{split}
\end{align}
and hence, the corresponding $U(1)$ gauge fluctuations maybe turned off by performing a compensating $U(1)$ gauge transformation generated by the gauge parameter
\begin{align}
    \begin{split}
        \Lambda^{\rm Sch.}_n(x)&=-\frac {%c_1
        \sqrt {3}\,Q} {4\sqrt{Q - a^2}}\,\frac{e^{in\theta}\tanh^{| n |}\left(\frac {\eta} {2}\right)}{\sqrt{|n|(n^2-1)}}\times\bigg[\,n\csch\eta \,(\csch\eta+| n |  \coth \eta )\,\bigg].
    \end{split}
\end{align}
The normalized Schwarzian modes obtained as pure diffeomorphism fluctuations from $\xi^{\rm Sch.}_n$ are given by,
\begin{align}\label{BMPVSCHM}
            \begin{split}
                h^{\rm (n)Sch.}_{\mu\nu}(x)\,dx^\mu dx^\nu%&=2\,\bar\nabla_{(A}\,\xi^{\rm Sch.}_n{}_{B)}\,dx^A\,dx^B\\
                &=\frac {1} {2\sqrt {2}\,\pi^{3/2}\big( {Q - a^2}\big)^{1/4}}\left[\frac{|n|(n^2-1)}{2}\right]^{1/2}\,\frac{(\sinh{\eta})^{|n|-2}}{(1+\cosh{\eta})^{|n|}}\,e^{in\theta}\\[5pt]
                \qquad\times&\left(d\eta^2+2i\,\text{sgn}(n)\,\sinh{\eta}\,d\eta\,d\theta-\sinh^2{\eta}\,d\theta^2\right),\quad n\in\mathbb{Z},\,|n|\geq2.\\[5pt]%\label{TZM_Explicit}
            \end{split}
        \end{align}
The modes associated with $n=0,\pm1$ are trivial and correspond to Killing isometries associated with the $AdS_2$ part of the near-horizon BMPV geometry. These fluctuations satisfy harmonic gauge condition.\\[5pt]
% The corrected eigenvalues for these modes take the following form,
% \begin{align}
%     \begin{split}
%         \kappa^{\rm (c)\,Sch.}_n&=\frac {3\,\pi\,|n|\,T} {2\sqrt {Q - a^2}}=\frac{3\,\pi\,|n|\,T}{2\sqrt{Q}}\,\sec{\omega},\qquad \big(a=-\sqrt{Q}\sin\omega\big)
%     \end{split}
% \end{align}
{\bf Rotational modes}\\[10pt]
Rotational modes are associated with pure diffeomorphism of the near-horizon BMPV background generated by the following set of local, non-normalizable vector fields,
\begin{align}
    \begin{split}\label{eq:rzm vec field}
        \xi^{\rm Rot.}_n&=%c_{\rm rot.}\big[
        \,H_{n}(x^{i})\,\partial_{\psi}+\gamma_n(x)\,\bar\nabla^{\mu}H_{n}(x^i)\,\partial_{\mu}%\big]
        \\[5pt]
        &=-\,%c_{\rm rot.}\,
        \frac{e^{in\theta}\tanh^{|n|}\left(\frac\eta2\right)}{\sqrt{2\pi|n|}(2Q-a^2)}\bigg[\,a\sqrt{Q-a^2}\,\big(i|n|\csch^2\eta\,\partial_\theta+n\csch\eta\,\partial_\eta\big)\\[5pt]
        &\qquad\qquad\qquad\qquad\qquad-\Big(2Q-a^2+\frac{a^2|n|}{1+\cosh\eta}\Big)\partial_\psi\,\bigg],%\qquad c_{\rm rot.}=\frac{\sqrt{2Q-a^2}}{\sqrt2\,\pi\,Q(Q-a^2)^{3/4}}
    \end{split}
\end{align}
where,
\begin{align}
    \begin{split}
        H_{n}(x^{i})&=\frac{1}{\sqrt{2\pi|n|}}\,\left(\frac{\sinh{\eta}}{1+\cosh{\eta}}\right)^{|n|}\,e^{in\theta}\,,\quad n\in\mathbb{Z},\,|n|\geq1,
    \end{split}
\end{align}
are non-normalizable scalars on $AdS_2$ defined in \eqref{eq:scalar_AdS2}, and
\begin{align}
    \begin{split}
        \gamma_n(x)&=-\,\text{sgn}(n)\,\frac{a\,Q\,\sqrt{Q-a^2}}{4\,(2Q-a^2)},
    \end{split}
\end{align}
is fixed from the harmonic gauge condition. Like the rotational modes of BTZ, here, the physical data about the fluctuations is encoded in the first term of $\xi_n^{\rm Rot.}$, whereas
the second piece is a small compensating gauge transformation needed to bring back the
generated metric fluctuations to the chosen gauge slice. For $n=0$, \eqref{eq:rzm vec field} gives back the rotational isometry generator $\partial_\psi$ in \eqref{eq:rot iso generator} up to an overall constant. Like the Schwarzian case, these vector fields do not Lie drag the extremal background field strength either, i.e.,
\begin{align}
    \begin{split}
        \mathcal{L}_{\xi^{\rm Rot.}_n}\,\bar{F}^{(0)}_{\mu\nu}&=0,
    \end{split}
\end{align}
and the $U(1)$ gauge modes obtained via Lie dragging the extremal background gauge field $\bar A_\mu^{(0)}$ along $\{\xi^{\rm Rot.}_n\}$ maybe compensated by a gauge transformation generated by the following gauge parameter,
\begin{align}
    \begin{split}
        \Lambda^{\rm Rot.}_n(x)&=-\,%c_{\rm rot.}\,
        \sqrt {\frac {3} {2\pi|n|}}\,\,\frac {a\,Q} {4\big (a^2 - 2 Q\big)}\,e^{in\theta}\tanh^{| n |}\left (\frac {\eta} {2}\right)\times\big (1+|n|\coth \eta\,\csch \eta\big).
    \end{split}
\end{align}
The normalized rotational modes obtained as pure diffeomorphism fluctuations from $\xi_n^{\rm Rot.}$ are found to be,
\begin{align}\label{BMPVROTM}
    \begin{split}
        h^{\rm (n) Rot.}_{\mu\nu}(x)\,dx^\mu dx^\nu&=\frac {\,e^{in\theta}\tanh^{| n |}\left (\frac {\eta} {2} \right)} {4\,\pi\sqrt{\pi|n|}\,\sqrt {2 Q - a^2}\,(Q - a^2)^{1/4}}\times\bigg[\,a\,n(| n | +\cosh \eta - 2)\,d\theta^2\\[5pt]
        &\qquad\quad-2\,i\,a|n|(|n|-1)\csch\,\eta\,d\theta d\eta+\,2\,i\,n\sqrt{Q-a^2}\,\cos{\hat\theta}\,d\theta d\phi\\[5pt]
        &\qquad\quad+2\,i\,n\sqrt{Q-a^2}\,d\theta d\psi-a\,\text{sgn}(n)\left(n^2-|n|\cosh\eta\right)\csch^2\eta\,\,d\eta^2\\[5pt]
        &\qquad\quad+| n |\sqrt{Q-a^2}\,\csch\,\eta\,\cos{\hat\theta}\,d\eta d\phi+| n |\sqrt{Q-a^2}\,\csch\,\eta\,\,d\eta\,d\psi\,\bigg]
    \end{split}
\end{align}
% obtained from the spontaneously broken $SU(2)_{\psi}$ sector of $SO(4)$, are generated by the following diffeo parameter,
% and the corrected eigenvalue is found to be,
% \begin{align}
%     \begin{split}
%         \kappa_n^{\rm(c)\,Rot.}&=\frac {|n|\,\pi\,T \,(1+3\cos 2\omega)\sec\omega} {\sqrt {Q}\,(3+\cos 2\omega)}
%     \end{split}
% \end{align}
{\bf SU(2) vector modes}\\[10pt]
% The generators $\{L_a\}$ of the unbroken $SU(2)_L$ sector of the original $SO(4)$ isometry group are,
% \begin{align}
%     \begin{split}
%         L_0&=\partial_\phi\,,\quad L_1=\frac{L_++L_-}{2}=-\sin{\phi}\,\partial_{\hat\theta}-\cot\hat\theta\,\cos\phi\,\partial_\phi+\csc\hat\theta\,\cos\phi\,\partial_{\psi}\,,\\[5pt]
%         &\text{and }\quad L_2=\frac{L_+-L_-}{2i}=\cos{\phi}\,\partial_{\hat\theta}-\cot\hat\theta\,\sin\phi\,\partial_\phi+\csc\hat\theta\,\sin\phi\,\partial_{\psi}
%     \end{split}
% \end{align}
% which generate the following $\mathfrak{su}(2)$ algebra,
% \begin{align}
%     \begin{split}
%         [L_a,L_b]&=\epsilon_{abc}\,L_c\,,\qquad\,a,b,c\in\{0,1,2\}
%     \end{split}
% \end{align}
% where, $\epsilon_{abc}$ is the fully anti-symmetric structure constant with
% \begin{align}
%     \epsilon_{210}=+1
% \end{align}
Due to the specific form of the angular momentum of the BMPV background, one of the $SU(2)$ symmetry survives. The diffeo generating vector field associated with these modes are given by,
\begin{align}
    \begin{split}
        \xi_{I}^{(n)}=\,\,H_n&(x^i)\,L_I\,+\,\gamma_I^{(n)}(x)\,\bar\nabla^\mu H_n(x^i)\,\partial_\mu,\qquad n\in\mathbb{Z},\,|n|>0,\\[5pt]
        \text{where, }\quad\gamma_I^{(n)}(x)=c_n&\hat f_I(x),\qquad c_n=-\,\text{sgn}(n)\,\frac{a\,Q\,\sqrt{Q-a^2}}{4\,(4Q-a^2)},\qquad I\in\{0,1,2\},%\quad\gamma_1^{(n)}(x)=c_n\sin\hat\theta\,\cos\phi\,,\quad\gamma_2^{(n)}(x)=c_n\sin\hat\theta\,\sin\phi\,,
        \\[5pt]
        \hat f_0(x)=&\cos\hat\theta\,,\quad \hat f_1(x)=\sin\hat\theta\,\cos\phi\,,\quad \hat f_2(x)=\sin\hat\theta\,\sin\phi,\label{eq:SU(2) mode diffeo param}
    \end{split}
\end{align}
are fixed by imposing the harmonic gauge condition. Here, $H_n(x^i)$ are $AdS_2$ scalars defined in \eqref{eq:scalar_AdS2} and $\{L_I\}$ are the $SU(2)_L$ generators in \eqref{SU2generator}, which can also be obtained by setting $n=0$ in \eqref{eq:SU(2) mode diffeo param}. The precise forms of the diffeo parameters are, 
\begin{align}
    \begin{split}
        \xi_0^{(n)}&=\frac {e^{in\theta}\tanh^{| n|}\left (\frac {\eta} {2} \right)} {\sqrt {2\pi|n|}}\bigg[-\frac{a\sqrt{Q-a^2}}{4Q-a^2}\,\cos\hat\theta\,\bigg(\frac{i|n|}{\sinh^2\eta}\partial_\theta+\frac n{\sinh\eta}\partial_\eta\bigg)\\[5pt]
        &\qquad\qquad\qquad\qquad+\,\partial_\phi+\frac{a^2}{4Q-a^2}\,|n|(\cosh\eta-1)\csch^2\eta\,\cos \hat\theta\,\partial_\psi\bigg],
    \end{split}
\end{align}
\begin{align}
    \begin{split}
        \xi_1^{(n)}&=\frac{ e^{in\theta}\, \tanh^{|n|}\!\left(\frac{\eta}{2}\right) }{ \sqrt{2\pi|n|} } \Bigg[-\frac{a\sqrt{Q-a^2}}{4Q-a^2}\,\sin\hat\theta\cos\phi\,\left(\frac{i\,|n|}{\sinh^2\eta}\partial_\theta+\frac n{\sinh\eta}\partial_\eta\right) -\sin\phi\,\partial_{\hat\theta} \\[2mm]
        &\qquad\qquad\qquad\qquad-\cos\phi\cot\hat\theta\,\partial_\phi+\cos\phi\left(\csc\hat\theta+\frac{a^2\,|n|\sin\hat\theta}{ (4Q-a^2)(1+\cosh\eta)}\right)\partial_\psi \Bigg],%-\frac {\sin \alpha e^{in\theta}\tanh^{| n |}\left (\frac {\eta} {2} \right)}{\sqrt{2\pi| n |}\left (a^2 - 4 Q \right)(\cosh \eta + 1)}\bigg [-2\left (a^2 - 
         %4 Q \right)\csc \alpha\cosh^2\left (\frac {\eta} {2} \right) (\cos \phi (\csc \alpha\, \partial_\psi - \cot \alpha\,\partial_\phi)\\[5pt]
         %&- \sin \phi\, \partial_\alpha) + \frac {1}{2} a\csch^2\left (\frac {\eta} {2} \right)\textit {sgn} (n)\cos \phi\left (-\partial_\eta\sqrt {Q - a^2} |n| \sinh (\eta) + 
         %n\left (-i\partial_\theta \sqrt {Q - a^2} + 
            %a(\cosh \eta - 1)\partial_\psi \right) \right) \bigg]
    \end{split}
\end{align}
and
\begin{align}
    \begin{split}
        \xi_2^{(n)}&=\frac{ e^{in\theta}\, \tanh^{|n|}\!\left(\frac{\eta}{2}\right) }{ \sqrt{2\pi|n|} } \Bigg[-\frac{a\sqrt{Q-a^2}}{4Q-a^2}\,\sin\hat\theta\sin\phi\,\left(\frac{i\,|n|}{\sinh^2\eta}\partial_\theta+\frac n{\sinh\eta}\partial_\eta\right) +\cos\phi\,\partial_{\hat\theta} \\[2mm]
        &\qquad\qquad\qquad\qquad-\sin\phi\cot\hat\theta\,\partial_\phi+\sin\phi\left(\csc\hat\theta+\frac{a^2\,|n|\sin\hat\theta}{ (4Q-a^2)(1+\cosh\eta)}\right)\partial_\psi \Bigg].
        %\frac {\csc\hat\theta e^{in\theta}\tanh^{| n |}\left (\frac {\eta} {2} \right)} {2\sqrt {2\pi|n|}\left (a^2 -4 Q \right) (1 + \cosh\eta)}\bigg[4\left (a^2- 4 Q \right)\cosh^2\left (\frac {\eta} {2} \right) (\partial_{\hat\theta}\sin\hat\theta\cos\phi + \sin \phi (\partial_\psi - \partial_\phi\cos\hat\theta))\\[5pt]
        %   &- a\sin^2\hat\theta\csch^2\left (\frac {\eta} {2} \right)\textit {sgn} (n)\sin\phi\left (-\partial_\eta\sqrt {Q - a^2} |% n |  \sinh\eta + 
        %  n\left (-i\partial_\theta\sqrt {Q - a^2} + 
        %a\partial_\psi\cosh \eta - a\partial_\psi\right)\right) \bigg]
    \end{split}
\end{align}
It can be checked that with the modification of the diffeo parameters with the small gauge term proportional to $\gamma_I^{(n)}(x)$, the metric fluctuations satisfy the required harmonic gauge condition. But the gauge fluctuation does not satisfy the Lorenz gauge condition and one needs to include a compensating $U(1)$ transformation to make it residual gauge invariant. However,
unlike the previous two cases, the extremal background field strength $\bar F_{\mu\nu}^{(0)}$ is now Lie-dragged non-trivially along $\{\xi^{(n)}_I\}$. Hence, the $U(1)$ gauge fluctuation modes cannot be turned off. They non-trivially couple with the metric fluctuations and appear in the quadratic action. The gauge parameter associated with such a compensating gauge transformation is,
%that enforces the $U(1)$ gauge modes obtained via Lie dragging of the extremal background gauge field $\bar A^{\,(0)}_B$ to be physical are,
\begin{align}
    \begin{split}
        \Lambda_I^{(n)}(x)&=-\xi_I\cdot\bar A^{(0)}-a\,\sqrt{\frac{3}{32\,\pi\,|n|}}\,e^{in\theta}\,\tanh^{|n|}\left(\frac{\eta}{2}\right)\hat f_I(x)\\[5pt] 
%     \end{split}
% \end{align}
% Explicitly these are
% \begin{align}
%     \Lambda_j^{(n)}(x)
        &=\sqrt {\frac {3|n|}{2\pi}}\,e^{in\theta}\,\tanh^{| n |}\left (\frac {\eta} {2}\right)\times\frac{ aQ\coth\eta\,\csch \eta} {4\left (4 Q - a^2 \right)}\,\hat f_I(x),\qquad I\in\{0,1,2\}.
    \end{split}
\end{align}
The $SU(2)$ vector zero modes of the quadratic kinetic operator spanning the metric and gauge fluctuations are thus, given by,
\begin{align}
    \begin{split}
        h_I^{(n)}{}_{\mu\nu}\,dx^\mu dx^\nu&=2\,\bar\nabla_{(\mu}\,\xi^{(n)}_I{}_{\nu)},\qquad
        a^{(n)}_I{}_\mu\,dx^\mu=\mathcal{L}_{\xi^{(n)}_I}\bar A^{(0)}_\mu\,dx^\mu+\bar\nabla_\mu\,\Lambda^{(n)}_I(x).
    \end{split}
\end{align}
For the uniformity of the manuscript, here we present explicit forms of the normalized $U(1)$ gauge and metric fluctuation modes.
\begin{align}\label{BMPVSU2A}
    \begin{split}
        a_I^{(n)}{}_\mu\,dx^\mu&=\mathcal N_0\,\frac{3aQ}{4(a^2-4Q)}\sqrt{\frac{3}{2\pi|n|}}\,e^{in\theta}\tanh^{|n|}\!\left(\frac{\eta}{2}\right)\big(|n|\,\csch\eta\,d\eta+in\,d\theta\big)\, f_I(x),\qquad I\in\{0,1,2\},
    \end{split}
\end{align}
\begin{align}
    \begin{split}
        h_0^{(n)}{}_{\mu\nu}\,dx^\mu dx^\nu&=\mathcal N_0\,\frac{Q\,e^{in\theta}\,\tanh^{|n|}\!\left(\frac{\eta}{2}\right)}{2\sqrt{2\pi|n|}\left(a^2-4Q\right)}\,\left(|n|\csch^2\eta\,d\eta+i\,n\csch\eta\,d\theta\right)\\[5pt]
        &\times\Bigg[\operatorname{sgn}(n)\,a\sqrt{Q-a^2}\left(|n|-\cosh\eta\right)\cos\hat\theta\,d\eta-\,4(Q-a^2)\sinh\eta \,\cos\hat\theta\,d\psi\\[5pt]
        &\qquad+\,i\,a\sqrt{Q-a^2}\left(|n|-4+3\cosh\eta\right)\sinh\eta\,\cos\hat\theta\,d\theta\\[5pt]
        &\qquad-\operatorname{sgn}(n)\,a\sqrt{Q-a^2}\,\sinh\eta\,\sin\hat\theta\,d\hat\theta+\sinh\eta\left(a^2-4Q+3a^2\cos^2\hat\theta\right)d\phi\Bigg],
        %\frac{ Q\,\sqrt{m}\, e^{in\theta}\, \csch^2\eta\, \tanh^m\!\left(\frac{\eta}{2}\right) }{ 4\sqrt{2\pi}\,(a^2-4Q)}\times \left( d\eta+i\,s\,\sinh\eta\,d\theta \right)\\[5pt]
        % &\times\Bigg[ 2as\Delta\cos\hat\theta \left( m\sinh\eta-\cosh\eta \right)d\eta+2ia\Delta\cos\hat\theta(m-1)\sinh\eta\,d\theta \\[5pt]
        % &+\left(5a^2-8Q+3a^2\cos2\hat\theta \right)d\phi+8(a^2-Q)\cos\hat\theta\,d\psi -as\Delta\sin\hat\theta\,d\hat\theta\Bigg],
    \end{split}
\end{align}
% where,
% \begin{align}
%     \begin{split}
%         m=|n|, \qquad s=\operatorname{sgn}(n), \qquad \Delta=\sqrt{Q-a^2}
%     \end{split}
% \end{align}
\begin{align}
    \begin{split}
        h_1^{(n)}{}_{\mu\nu}&\,dx^\mu dx^\nu=\mathcal N_0\,\frac{Q\,e^{in\theta}\tanh^{|n|}\!\left(\frac{\eta}{2}\right)}{2\sqrt{2\pi|n|}\,(a^2-4Q)}\left(|n|\csch^2\eta\, d\eta+i\,n\cosh\eta\,d\theta\right)\\[5pt]
        &\quad\times\Bigg[\operatorname{sgn}(n)\,a\sqrt{Q-a^2}\,\big(|n|-\cosh\eta\big)\,\sin\hat\theta\,\cos\phi\,d\eta+4(a^2-Q)\sinh\eta\,\sin\hat\theta\,\cos\phi\,d\psi\\[5pt]
        &\qquad+\,i\,a\sqrt{Q-a^2}\big(|n|-4+3\cosh\eta\big)\,\sinh\eta\,\,\sin\hat\theta\,\cos\phi\,d\theta\\[5pt]
        &\qquad+\sinh\eta\,\Big(\operatorname{sgn}(n)\,a\sqrt{Q-a^2}\,\cos\hat\theta\,\cos\phi-(a^2-4Q)\sin\phi\Big)\,d\hat\theta\\[5pt]
        &\qquad+a\sinh\eta\,\sin\hat\theta\,\Big(3a\cos\hat\theta\cos\phi-\operatorname{sgn}(n)\sqrt{Q-a^2}\,\sin\phi\Big)\,d\phi\Bigg],
        %(Form-2) \frac{Q\,e^{in\theta}\tanh^{|n|}\!\left(\frac{\eta}{2}\right)}{\sqrt{2\pi|n|}(a^2-4Q)}\,\Bigg[2\,a\sqrt{Q-a^2}\cos\hat\theta\,\operatorname{sgn}(n)\left(|n|-\cosh\eta\right)d\eta\\[5pt]
        % &+2ia\sqrt{Q-a^2}\left(|n|+3\cosh\eta-4\right)\sinh\eta\,\cos\hat\theta\,d\theta-16(Q-a^2)\sinh\eta\sin\hat\theta\cos\phi\,d\psi\\[5pt]
        % &+\sinh\eta\bigg(3a^2\sin 2\hat\theta\,\cos\phi-2\operatorname{sgn}(n)a\sqrt{Q-a^2}\sin\hat\theta\sin\phi\bigg)d\phi-2\big(\operatorname{sgn}(n)a\sqrt{Q-a^2}\cos\hat\theta\cos\phi\\[5pt]
        % &+(a^2-4Q)\sin\phi\big)\,d\hat\theta\Bigg]\times\left(|n|\csch^2\eta\,d\eta+i\,n\csch\eta\,d\theta\right)
        %(form-1)\mathcal{N}_m(\eta)\, e^{in\theta} \left( d\eta+i\,s\sinh\eta\,d\theta \right) \Omega_s , \qquad\mathcal{N}_m(\eta) =\frac{ Q\sqrt{m}\, \csch^2\eta\, \tanh^{m-2}\!\left(\frac{\eta}{2}\right) }{ 2\sqrt{2\pi}\, (a^2-4Q)\, (1+\cosh\eta)^2 }\\[10pt]
        % &\qquad\Omega_s= a\Delta\cos\phi\sin\alpha \left( sm\sinh\eta-\cosh\eta \right)d\eta-ia\Delta\cos\phi\sin\alpha \left( 4+sm+3s\cosh\eta \right) \sinh\eta\,d\theta \\
        % &\qquad\qquad+4(a^2-Q)\cos\phi\sin\alpha\, \sinh\eta\,d\psi+a\cos\alpha\cos\phi \left( s\Delta\,d\alpha +3a\sin\alpha\,d\phi \right) \sinh\eta \\
        % &\qquad\qquad-\sin\phi \left[ (a^2-4Q)d\alpha +a\Delta\sin\alpha\,d\phi \right] \sinh\eta
    \end{split}
\end{align}
\begin{align*}
    \begin{split}
        h_2^{(n)}{}_{\mu\nu}&\,dx^\mu dx^\nu=\mathcal N_0\,\frac{Q\,e^{in\theta}\tanh^{|n|}\!\left(\frac{\eta}{2}\right)}{2\sqrt{2\pi|n|}\,(a^2-4Q)}\left(|n|\csch^2\eta\, d\eta+i\,n\cosh\eta\,d\theta\right)\\[5pt]
        &\quad\times\Bigg[\operatorname{sgn}(n)\,a\sqrt{Q-a^2}\,\big(|n|-\cosh\eta\big)\,\sin\hat\theta\,\sin\phi\,d\eta+4(a^2-Q)\sinh\eta\,\sin\hat\theta\,\sin\phi\,d\psi\\[5pt]
        &\qquad\quad\,\,\,\,+\,i\,a\sqrt{Q-a^2}\big(|n|-4+3\cosh\eta\big)\,\sinh\eta\,\,\sin\hat\theta\,\sin\phi\,d\theta
    \end{split}
\end{align*}
\begin{align}\label{BMPVSU2M}
    \begin{split}
        &+\sinh\eta\,\Big(\operatorname{sgn}(n)\,a\sqrt{Q-a^2}\,\cos\hat\theta\,\sin\phi+(a^2-4Q)\cos\phi\Big)\,d\hat\theta\\[5pt]
        &+a\sinh\eta\,\sin\hat\theta\,\Big(3a\cos\hat\theta\sin\phi+\operatorname{sgn}(n)\sqrt{Q-a^2}\,\cos\phi\Big)\,d\phi\Bigg],
        %(form-1)\frac{Q\sqrt{|n|}\,e^{in\theta}\tanh^{|n|}\!\left(\frac{\eta}{2}\right)}{2\sqrt{2\pi}\,(a^2-4Q)}\times\left(\csch^2\eta\,d\eta+i\,\operatorname{sgn}(n)\csch\eta\,d\theta\right)\\[5pt]
        % &\qquad\times\Bigg[a\sqrt{Q-a^2}\,\sin\hat\theta\sin\phi\,\operatorname{sgn}(n)\left(|n|-\cosh\eta\right)d\eta+(a^2-4Q)\cos\phi\,d\hat\theta\\[5pt]
        % &\qquad+a\sqrt{Q-a^2}\,\sin\hat\theta\cos\phi\,\operatorname{sgn}(n)\sinh\eta\,d\phi+3a^2\sin\hat\theta\cos\hat\theta\sin\phi\,\sinh\eta\,d\phi\\[5pt]
        % &\qquad+4(a^2-Q)\,\operatorname{sgn}(n)\sin\hat\theta\sin\phi\,\sinh\eta\,d\psi+a\sqrt{Q-a^2}\,\sin\hat\theta\cos\hat\theta\sin\phi\,\sinh\eta\,d\hat\theta\\[5pt]
        % &\qquad
        % +i\,a\sqrt{Q-a^2}\,\sin\hat\theta\sin\phi\,\left(|n|-4-3\cosh\eta\right)\sinh\eta\,d\theta\Bigg]
        %(form-2)\frac{ Q\sqrt{m}\, e^{in\theta}\, \csch^2(\eta)\, \tanh^m\!\left(\frac{\eta}{2}\right) }{ 2\sqrt{2\pi}\, (a^2-4Q) } \,\Xi_s\\[10pt]
        % \Xi_s={}& as\Delta\sin\alpha\sin\phi \left( m-\cosh\eta \right)d\eta \\[1mm] &+ \sinh\eta \Bigg[ \left( (a^2-4Q)\cos\phi + as\Delta\sin\phi\cos\alpha \right)d\alpha \\ &\qquad\qquad +4(a^2-Q)\sin\phi\,d\psi \\ &\qquad\qquad +\left( as\Delta\cos\phi + 3a^2\sin\phi\sin\alpha\cos\alpha \right)d\phi \\ &\qquad\qquad +i\,a\Delta\sin\phi \left( m-4+3\cosh\eta \right)d\theta \Bigg]
    \end{split}
\end{align}
where,
\begin{align}
    \mathcal{N}_0&=\frac {\pi^2 Q^2\sqrt {Q - a^2}\left (16 a^4 - 87 a^2 Q + 
     128 Q^2 \right)} {8\left (a^2 - 4 Q \right)^2},
\end{align}
is the normalization factor commonly shared by modes across the entire $SU(2)_L$ sector. Since the metric and $U(1)$ gauge fluctuation modes can not be decoupled in this case, we define our fluctuation mode vectors as $X_n=(h_{\mu\nu}^{(n)},a_\mu^{(n)})^T$ and fix the orthonormality relation as follows,
\begin{align}
    \langle X_m|X_n\rangle&=\int d^5x\,\sqrt{\bar g}%{}^{(0)}
    \left[h^{\mu\nu}_{(m)}h^{(n)}_{\mu\nu}+a^\mu_{(m)} a^{(n)}_\mu\right]=\delta_{m+n,0}.
\end{align}
% where,
% \begin{align}
%     m=|n|, \qquad s=\operatorname{sgn}(n), \qquad \Delta=\sqrt{Q-a^2}
% \end{align}
% and the corrected eigenvalue for each of these different set of modes is found to be same and is given by,
% \begin{align}
%     \begin{split}
%         \kappa^a_n{}^{(c)}&=\frac {8\,\pi\,|n|\,T\sqrt {Q - a^2}\left (4 a^4 - 33 a^2 Q + 
%       48 Q^2 \right)} {3\,Q\left (16 a^4 - 87 a^2 Q + 128 Q^2 \right)}\\[5pt]
%       &=\frac {8\,\pi\,| n |\, 
%    T\cos \omega\,(29\cos 2\omega + \cos 4\omega + 
%       66)} {3\sqrt {Q}\,(71\cos 2\omega + 4\cos 4\omega + 181)}
%     \end{split}
% \end{align}
% Notice that unlike the rotational modes the above correction is positive for any $\omega\in\big[0,\frac{\pi}{2}\big)$.\\
{\bf U(1) gauge zero modes}\\[10pt]
% The corrected eigenvalue takes the following form,
% \begin{align}
%     \begin{split}
%         \kappa^{(c)\,U(1)}_n&=\frac {6\,\pi\,T}{\sqrt {Q - 
%       a^2}\left (3 | n | +n^2 + 2 \right)}\,\Bigg[| 
%       n |  \left (3 n^2\cosh \eta_ 0 + 7 n^2 + 2 \right)\\& + 
%        n^2\left (\left (2 + n^2 \right)\cosh \eta_ 0 + 2 n^2 + 
%           7 \right) \Bigg]\times\sech^4\left (\frac {\eta _ 0} {2} \right)\tanh^{2 | n |}\left (\frac {\eta_ 0} {2} \right) 
%     \end{split}
% \end{align}
% However, in the limit $\eta_0\rightarrow\infty$ we get,
% \begin{align}
%     \begin{split}
%         \kappa^{(c),\,U(1)}_n&=\frac {48\,\pi\,n^2\,T} {\sqrt {Q - a^2}}\,e^{-\eta_0}\sim0
%     \end{split}
% \end{align}
% Therefore, these modes do not contribute to log $T$ corrections and do not get uplifted in the near-extremal background.
Finally the $U(1)$ gauge zero modes are an uplift of the vector modes in $AdS_2$ to 5d and are given by,
\begin{align}
    \begin{split}
        a^{(n)}_\mu{}^{U(1)}\,dx^\mu&=\bar\nabla_\mu H_n(x^i)\,dx^\mu,\qquad n\in\mathbb{Z},|n|\geq1\\[5pt]
        &=\frac {1}{\sqrt {2\pi|n|}}\,e^{in\theta}\tanh^{| n |}\left (\frac {\eta} {2} \right)\left (| n | \csch\eta\, d\eta + i  n\, d\theta\right),
    \end{split}
\end{align}
where, $H_n(x^i)$ is the same scalar function defined in \eqref{eq:scalar_AdS2}. However, the above discrete gauge modes are normalizable in the BMPV background.
\subsection{Quadratic action}

Since the $SU(2)$ modes contain both the metric and gauge fluctuations, we need to study the full quadratic action, including the mixed metric and gauge fluctuation terms. Here we present the complete quadratic action governing the near-extremal BMPV system.  
Up to an overall factor of $-\frac{1}{16\pi G_5}$, the quadratic variations of the different pieces in the action \eqref{5d_action} post Euclidean continuation are as follows,
\begin{align}
    \begin{split}
        \delta^2(\sqrt g \,R)&=\frac{\sqrt {\bar g}}{4}\bigg[\,h_{\mu\nu}\bar\nabla_\rho \bar\nabla^\rho h^{\mu\nu}-h\bar\nabla_\rho \bar\nabla^\rho h-2h^{\nu\rho}\bar\nabla_\mu \bar\nabla_\nu h^\mu{}_\rho+2h^{\mu\nu}\bar\nabla_\mu \bar\nabla_\nu h\\[5pt]
        &\qquad+2\bar R_{\mu\nu}\big(2h^{\mu\rho}h^\nu{}_\rho-h \,h^{\mu\nu}\big)-\bar R\big(h_{\mu\nu}h^{\mu\nu}-\frac12 h^2\big)\,\bigg],\\[5pt]
        -\delta^2(\sqrt g \,F^2)&=\frac{\sqrt{\bar g}}{4}\bigg[-4f_{\mu\nu}f^{\mu\nu}-4\bar F_{\mu\nu}\bar F_{\rho\sigma}h^{\mu\rho}h^{\nu\sigma}-8\bar F_{\mu\sigma}\bar F_\nu{}^\sigma h^{\mu\rho}h^\nu{}_\rho+4\bar F_{\mu\nu}\bar F_\rho{}^\nu\,h \,h^{\mu\rho}\\[5pt]
        &\qquad+\bar F^2 h_{\mu\nu}h^{\mu\nu}-\frac12\bar F^2 h^2+16\bar F_{\mu\nu}h^{\mu\rho}f_\rho{}^\nu-4\bar F_{\mu\nu}\,f^{\mu\nu}h\bigg],\\[5pt]
        \delta^2({\rm CS-term})&=-\frac{2}{3\sqrt 3}\,\epsilon^{\mu\nu\rho\sigma\lambda}\big(2 \,a_\mu f_{\nu\rho}\bar F_{\sigma\lambda}+\bar A_\mu f_{\nu\rho}f_{\sigma\lambda}\big),
    \end{split}
\end{align}
where, $\bar F^2=\bar F_{\rho\sigma}\bar F^{\rho\sigma}$ and $h=h^\rho{}_\rho$. To gauge fix the theory we add the following gauge-fixing terms to the action at quadratic order,
\begin{align}
    \begin{split}
        -\frac{\sqrt{\bar g}}{2}(\bar\nabla^\rho h_{\rho\mu}-\frac12\bar\nabla_\mu h)(\bar\nabla_\sigma h^{\sigma\mu}-\frac12\bar\nabla^\mu h)-2\sqrt{\bar g}(\bar\nabla_\mu a^\mu)^2,
    \end{split}
\end{align}
which up to total derivatives takes the following form,
\begin{align}
    \begin{split}
        \sqrt{\bar g}\,\bigg[\,\frac{1}{2}h_{\mu\nu}\bar\nabla^\mu \bar\nabla^\rho h_\rho{}^\nu-\frac12h_{\mu\nu}\bar\nabla^\mu \bar\nabla^\nu h+\frac18h\bar\nabla_\rho \bar\nabla^\rho h+2\,a_\mu \bar\nabla^\mu \bar\nabla^\nu\,a_\nu\,\bigg].
    \end{split}
\end{align}
Discarding total derivatives and imposing the EOMs, the full gauge-fixed quadratic action is given by,
\begin{align}
    \begin{split}
        S^{(2)}&=S^{(2)}_{hh}+S^{(2)}_{ha}+S^{(2)}_{aa},
    \end{split}
\end{align}
where,
\begin{align}
    \begin{split}
        &S^{(2)}_{hh}=-\frac{1}{16\pi G_5}\int d^5x\,\,\frac{\sqrt {\bar g}}{4}\bigg[\,h_{\mu\nu}\bar\nabla_\rho \bar\nabla^\rho h^{\mu\nu}-\frac12 h\bar\nabla_\rho \bar\nabla^\rho h+2\bar R^{\lambda}{}_{\rho\mu\nu}h^{\nu\rho} h^\mu{}_\lambda\\[5pt]
        &\qquad\qquad\qquad\qquad+\frac{1}{3}\bar F^2h^2-4\bar F_{\mu\nu}\bar F_{\rho\sigma}h^{\mu\rho}h^{\nu\sigma}-4\bar F_{\mu\sigma}\bar F_\nu{}^\sigma h^{\mu\rho}h^\nu{}_\rho\,\bigg]
    \end{split}
\end{align}
\begin{align}
        S^{(2)}_{ha}&=-\frac{1}{16\pi G_5}\int d^5x\,\,\sqrt{\bar g}\,\bigg[\,4\bar F_{\mu\nu}h^{\mu\rho}f_\rho{}^\nu-\bar F_{\mu\nu}h\,f^{\mu\nu}\,\bigg],\\[10pt]
        \text{and}\qquad S^{(2)}_{aa}&=-\frac{1}{16\pi G_5}\int d^5x\bigg[\,\sqrt{\bar g}\,\big(\,2\,a_\mu \bar\nabla_\rho \bar\nabla^\rho a^\mu-2\,a_\mu \bar R^{\mu\nu}a_\nu\big)-\frac{2}{\sqrt 3}\,\epsilon^{\mu\nu\rho\sigma\lambda}\,\,a_\mu \bar F_{\nu\rho}f_{\sigma\lambda}\,\bigg].
\end{align}
\subsection {Near-extremal log corrections }\label{sub-sec:log T corrections near-BMPV}
\subsubsection*{Schwarzian modes}
The corrected eigenvalues for these modes take the following form,
\begin{align}
    \begin{split}
        \lambda^{\rm \,Sch.}_n&=\frac {3\,\pi\,|n|\,T} {2\sqrt {Q - a^2}}=\frac{3\,\pi\,|n|\,T}{2\sqrt{Q}}\,\sec{\omega},\qquad \big(a=-\sqrt{Q}\sin\omega\big)
    \end{split}
\end{align}
\begin{figure}[htbp]
    \centering % Centers the image horizontally
    \includegraphics[width=0.6\textwidth]{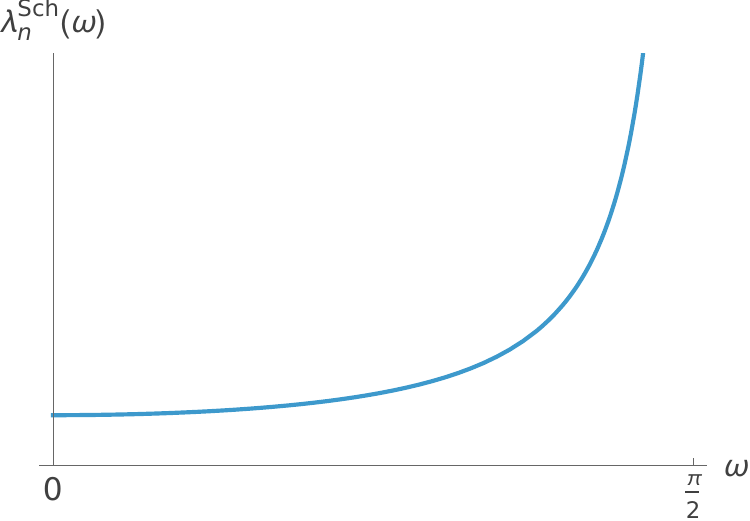} % Scales image to 70% of text width
    \caption{a schematic plot of $\lambda^{\rm Sch.}_n(\omega)\text{ vs }\omega$}
    \label{fig:eigen_val_Sch} % Used to cross-reference with \ref
\end{figure}\\[5pt]
Thus, the near-extremal corrected quadratic action is,
\begin{align}
    \begin{split}
        S^{(2)}&\sim\sum_{|n|\geq2} c_{-n}\, c_n\times\frac{1}{16\pi G_5}\times\lambda^{\rm Sch.}_n\\[5pt]
        &\sim\sum_{|n|\geq2}\frac{32}3\,QG_5\times\frac{\tilde c_{-n}\,\tilde c_n}{16\pi G_5}\times\lambda^{\rm Sch.}_n,\qquad\tilde c_{\pm n}\in\mathbb{C}
    \end{split}
\end{align}
and subsequently, the corrected logarithm of the partition function is given by,
\begin{align}
    \begin{split}
        \log Z&\sim-\frac12\times\sum_{|n|\geq2}\log\left(\frac{|n|\sqrt Q\,T}{\cos\omega}\right)
    \end{split}
\end{align}
%\textcolor{red}{(could we absorb the numerical factor $3/32$ into $\tilde c_{-n}\,\tilde c_n$ ?)}  
regularizing which we get,
\begin{align}\label{BMPVSCH}
    \begin{split}
        \log Z^{\rm reg.}&=\frac32\log\left(\frac{T}{T_{\rm Sch.}}\right),\qquad T_{\rm Sch.}=%\frac{32\sqrt{Q-a^2}}{3Q}=\frac{32\cos\omega}{3\sqrt Q}\,\textcolor{red}{\overset{\text{OR}}{=}
        \frac{\cos\omega}{\sqrt Q}
    \end{split}
\end{align}
\subsubsection*{Rotational modes}
The corrected eigenvalue corresponding to the rotational modes is found to be,
\begin{align}
    \begin{split}
        \lambda^{\rm Rot.}_n&=\frac {|n|\,\pi\,T \,(1+3\cos 2\omega)\sec\omega} {\sqrt {Q}\,(3+\cos 2\omega)}\\[5pt]
        &=\begin{cases}
            \,>\,0\,,&\,\,0\leq\omega<\omega_0=\dfrac12\cos^{-1}\left(-\dfrac13\right)\,,\,\,\forall\,\,|n|\geq1\\[10pt]
            \,<\,0\,,&\,\,\dfrac\pi2>\omega>\omega_0=\dfrac12\cos^{-1}\left(-\dfrac13\right)\,,\,\,\forall\,\,|n|\geq1
        \end{cases}
    \end{split}
\end{align}
\begin{figure}[htbp]
    \centering % Centers the image horizontally
    \includegraphics[width=0.6\textwidth]{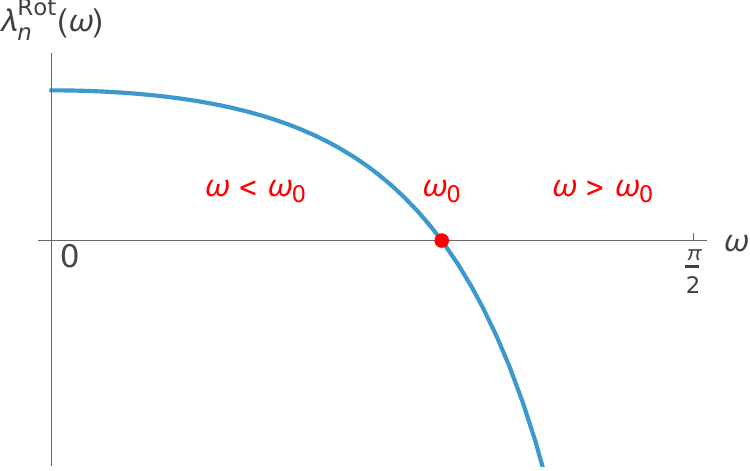} % Scales image to 70% of text width
    \caption{a schematic plot of $\lambda_n^{\rm Rot.}(\omega)\text{ vs }\omega$; $\omega_0=\frac12\cos^{-1}(-\frac13)$.}
    \label{fig:Eigen_Value_Rot} % Used to cross-reference with \ref
\end{figure}\\[5pt]
Thus, the near-extremal corrected quadratic action is,
\begin{align}
    \begin{split}
        S^{(2)}&\sim\sum_{|n|\geq1} c_{-n}\,c_n\times\frac{1}{16\pi G_5}\times\lambda^{\rm Rot.}_n\\[10pt]
        %S^{(2)}
        &\sim\begin{cases}
            \sum\limits_{|n|\geq1}16\,QG_5\times\dfrac{\tilde c_{-n}\,\tilde c_n}{16\pi G_5}\times\lambda^{\rm Rot.}_n,&0\leq\omega<\omega_0%,\forall\,|n|\geq1
            \\[10pt]
            \sum\limits_{|n|\geq1}-16\,QG_5\times\dfrac{\tilde c_{-n}\,\tilde c_n}{16\pi G_5}\times\lambda^{\rm Rot.}_n,&\omega_0<\omega<\dfrac\pi2%,\forall\,|n|\geq1
        \end{cases}\,\,,\qquad\tilde c_{\pm n}\in\mathbb{C}.
    \end{split}
\end{align}
The above result is a distinct feature of the rotational zero modes, as compared to the Schwarzian sectors modes discussed last and the $SU(2)$ or the $U(1)$ modes that we present next. The rotational modes contribution is zero at $\omega_0= \frac12\cos^{-1}\left(-\frac13\right)$ and changes sign across this value. This clearly indicates that depending on the saddle, the fluctuations $\tilde c_n$ would get contribution from a particular contour. Incorporating this subsequently, the corrected logarithm of the partition function is given by,
\begin{align}
    \begin{split}
        \log Z&\sim%-\frac12\sum_{|n|\geq1}\log\left(\frac{|n|\,Q\,T}{16\sqrt{Q-a^2}}\cdot\left|\frac{2Q-3a^2}{2Q-a^2}\right|\right)\textcolor{red}{\overset{?}{\sim}
        -\frac12\sum_{|n|\geq1}\log\left(\dfrac{|n|\,\sqrt Q\,T}{\cos\omega}\cdot\left|\dfrac{1+3\cos2\omega}{3+\cos2\omega}\right|\right)
    \end{split}
\end{align}
regularizing\footnote{From zeta regularization we have $\sum_{n=1}^{\infty}\log \left(\alpha\,n^k\right)=\zeta(0)\log\alpha-k\cdot\zeta'(0)$, where, $\zeta(0)=-\frac12$, and $\zeta'(0)=-\frac12\,\log2\pi$.} which we get,
\begin{align}\label{BMPVROT}
    \begin{split}
        \log Z^{\rm reg.}&=\frac12\log\left(\frac{T}{T_{\rm Rot.}}\right),\qquad
        T_{\rm Rot.}=%\frac{16\sqrt{Q-a^2}}{Q}\cdot\left|\frac{2Q-a^2}{2Q-3a^2}\right|=\frac{16\sec\omega}{\sqrt Q}\left|\frac{3+\cos2\omega}{1+3\cos2\omega}\right|\textcolor{red}{\overset{?}{=}
        \frac{\cos\omega}{\sqrt Q}\left|\frac{3+\cos2\omega}{1+3\cos2\omega}\right|
    \end{split}
\end{align}
The above result is valid for $\omega\in\left[0,\omega_0\right)\cup\left(\omega_0,\frac\pi2\right)$, where, $\omega_0=\frac12\cos^{-1}\left(-\frac13\right)$. In particular, the $\omega=\omega_0$ mode does not get uplifted as the corresponding eigen value correction remains zero. Hence they do not contribute to the $\log \, T$ term.
\subsubsection*{SU(2) vector modes}
% The generators $\{L_a\}$ of the unbroken $SU(2)_\phi$ symmetry are as follows,
% \begin{align}
%     \begin{split}
%         L_0&=\partial_\phi\,,\quad L_1=\frac{L_++L_-}{2}=-\sin{\phi}\,\partial_{\hat\theta}-\cot\hat\theta\,\cos\phi\,\partial_\phi+\csc\hat\theta\,\cos\phi\,\partial_{\psi}\,,\\[5pt]
%         &\text{and }\quad L_2=\frac{L_+-L_-}{2i}=\cos{\phi}\,\partial_{\hat\theta}-\cot\hat\theta\,\sin\phi\,\partial_\phi+\csc\hat\theta\,\sin\phi\,\partial_{\psi}
%     \end{split}
% \end{align}
% The diffeo generating vector field associated with these modes are given by,
% \begin{align}
%     \begin{split}
%         \xi_{a}^{(n)}&=H_n(x^\mu)\,L_a\,+\,\gamma_a^{(n)}(x)\,\nabla^AH_n(x^\mu)\,\partial_A\\[5pt]
%         \text{where, }\quad\gamma_0^{(n)}(x)=c_n&\cos\hat\theta\,,\quad\gamma_1^{(n)}(x)=c_n\sin\hat\theta\,\cos\phi\,,\quad\gamma_2^{(n)}(x)=c_n\sin\hat\theta\,\sin\phi\,,\\[5pt]
%         \text{and }\quad c_n&=-\,\text{sgn}(n)\,\frac{a\,Q\,\sqrt{Q-a^2}}{4\,(4Q-a^2)}
%     \end{split}
% \end{align}
% and the gauge parameter associated with the compensating gauge transformation that enforces the $U(1)$ gauge modes obtained via Lie dragging of the extremal background gauge field $\bar A^{\,(0)}_B$ to be physical are,
% \begin{align}
%     \begin{split}
%         \Lambda_a^{(n)}(x)&=-\xi_a\cdot\bar A^{(0)}-a\,\sqrt{\frac{3}{32\,\pi\,|n|}}\,e^{in\theta}\,\tanh^{|n|}\left(\frac{\eta}{2}\right)f_a(x)\\[5pt]
%         \text{where, }\quad f_0(x)&=\cos\hat\theta\,,\quad f_1(x)=\sin\hat\theta\,\cos\phi\,,\quad f_2(x)=\sin\hat\theta\,\sin\phi\
%     \end{split}
% \end{align}
The corrected eigenvalue for each of these different set of $SU(2)$ vector modes is found to be same and is given by,
\begin{align}
    \begin{split}
        \lambda^{I}_n&=\frac {8\,\pi\,|n|\,T\sqrt {Q - a^2}\left (4 a^4 - 33 a^2 Q + 
      48 Q^2 \right)} {3\,Q\left (16 a^4 - 87 a^2 Q + 128 Q^2 \right)}\\[5pt]
      &=\frac {8\,\pi\,| n |\, 
   T\cos \omega\,(29\cos 2\omega + \cos 4\omega + 
      66)} {3\sqrt {Q}\,(71\cos 2\omega + 4\cos 4\omega + 181)}
    \end{split}
\end{align}
\begin{figure}[htbp]
    \centering % Centers the image horizontally
    \includegraphics[width=0.6\textwidth]{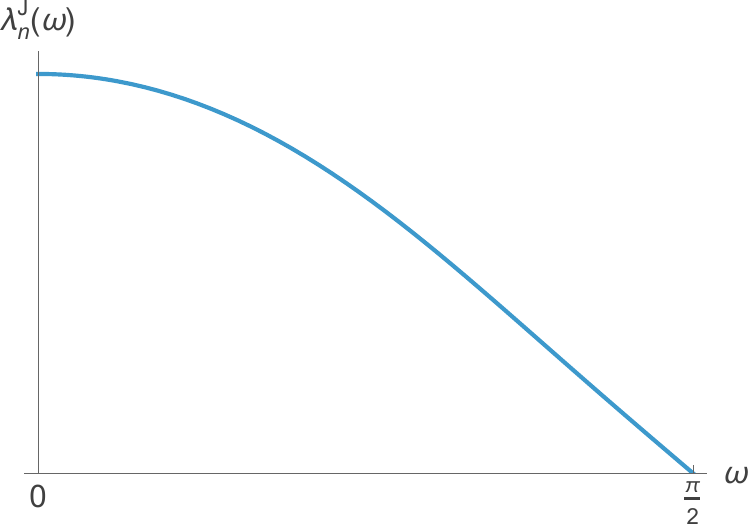} % Scales image to 70% of text width
    \caption{a schematic plot of $\lambda_n^{J}(\omega)\text{ vs }\omega$, $J\in\{0,1,2\}$.}
    \label{fig:eigen_val_SU(2)} % Used to cross-reference with \ref
\end{figure}\\[5pt]
Unlike the rotational modes the above correction is positive for any $\omega\in\big[0,\frac{\pi}{2}\big)$. The near-extremal corrected quadratic action for each of these $SU(2)$ modes is
\begin{align}
    \begin{split}
        S^{(2)}_I&\sim\sum_{|n|\geq1} c_{-n}\,c_n\times\frac{1}{16\pi G_5}\times\lambda_n^{I},\qquad I\in\{0,1,2\},\\[5pt]
        &\sim\sum_{|n|\geq1}6\,QG_5\times\frac{\tilde c_{-n}\,\tilde c_n}{16\pi G_5}\times\lambda_n^{I},\qquad\tilde c_{\pm n}\in\mathbb{C}.
    \end{split}
\end{align}
Hence, the logarithm of the partition function with near-extremal corrections is
\begin{align}
    \begin{split}
        \log Z&\sim-\sum_{I%\in\{0,1,2\}
        }\,\frac12\times\sum_{|n|\geq1}\log\left(\dfrac{|n|\sqrt {Q}\,T\,(66+\cos 4\omega +29\cos 2\omega)} {\sec\omega\,(181+71\cos 2\omega + 4\cos 4\omega)}\right),
    \end{split}
\end{align}
regularizing which we get,
\begin{align}\label{BMPVSU2}
    \begin{split}
        \log Z^{\rm reg.}&=\frac32\,\log\left(\frac{T}{T_{ SU(2)}}\right),
    \end{split}
\end{align}
where,
\begin{align}
    \begin{split}
        T_{SU(2)}&=\frac {\,(181+71\cos 2\omega + 4\cos 4\omega )\sec\omega} {\sqrt {Q} \,(66+\cos 4\omega +29\cos 2\omega)}.
    \end{split}
\end{align}
\subsubsection*{U(1) gauge zero modes}
The corrected eigenvalue takes the following form,
\begin{align}
    \begin{split}
        \lambda^{U(1)}_n&=\frac {6\,\pi\,T}{\sqrt {Q - 
      a^2}\left (3 | n | +n^2 + 2 \right)}\,\Bigg[| 
      n |  \left (3 n^2\cosh \eta_ 0 + 7 n^2 + 2 \right)\\& + 
       n^2\left (\left (2 + n^2 \right)\cosh \eta_ 0 + 2 n^2 + 
          7 \right) \Bigg]\times\sech^4\left (\frac {\eta _ 0} {2} \right)\tanh^{2 | n |}\left (\frac {\eta_ 0} {2} \right) 
    \end{split}
\end{align}
where, $\eta_0$ is the $AdS_2$ radial cut-off. However, in the limit $\eta_0\rightarrow\infty$ we get,
\begin{align}
    \begin{split}
        \lambda^{U(1)}_n&=\frac {48\,\pi\,n^2\,T} {\sqrt {Q - a^2}}\,e^{-\eta_0}\sim0
    \end{split}
\end{align}
Therefore, these modes do not contribute to log $T$ corrections and do not get uplifted in the near-extremal background.

\subsection{Scaling features}
Finally we present the scaling feature of our answer. Consider the size of the BMPV black hole to scale as $r_0$. We fix all charges to their BMPV values which scale uniformly with $r_0$, while we introduce a small temperature $T$ in the near-extremal extension. Thus, we consider these two independent length scales, $r_0$ and $T$. The scaling of the charges are,
\begin{align}
    Q \sim r_0^2, \quad J \sim r_0^3
\end{align}
This gives,
\begin{align}
    a \sim r_0, \quad \omega \sim \mathcal{O}(1)
\end{align}
Thus we can now find the scaling of the arguments of the logarithms in the slightly non-zero sectors:
\begin{align}
    T_{\rm Sch.} &= \frac{\cos\omega}{\sqrt Q} \sim r_0^{-1} \\[5pt]
    T_{SU(2)}&=  \frac {\,(181+71\cos 2\omega + 4\cos 4\omega 
     )\sec\omega} {\sqrt {Q} \,(66+\cos 4\omega +29\cos 2\omega)} \sim r_0^{-1} \\[5pt]
     T_{\rm Rot.}&= \frac{\cos\omega}{\sqrt Q}\left|\frac{3+\cos2\omega}{1+3\cos2\omega}\right| \sim r_0^{-1}
\end{align}
We find that, similar to the near-extremal BTZ, the near-extremally extended BMPV also gets a particular combination of the scaling parameters inside the log terms. All the arguments of the logarithms are $\log (r_0 T)$.

\section{Results and Discussions}\label{section5}
In this section, we first summarize the main results that we have obtained. We have studied the rotating black hole solutions in three, four and five spacetime dimensions. While the results of three dimensional BTZ and four dimensional Kerr-like solutions are already known, we present our understanding about these systems. For the five dimensional minimal BMPV system, our results are fundamental. The main results obtained are enumerated below :

\begin{itemize}
\item {\bf BTZ :} We have found the extremal contributions to the quantum log term in the BTZ entropy, by treating the horizon radius and the cosmological constants as two independent scale parameters. In the near-extremal extension, we have found the log temperature corrections. The results are given in equations \eqref{BTZER1}-\eqref{BTZER3} and \eqref{BTZNER}.
\item {\bf Kerr :} We agree with the existing literature. The rotational zero modes require further investigation. 
\item {\bf BMPV :} The $\log$ correction for BMPV solutions were already known in the literature. We have explicitly constructed the extremal zero modes in equations \eqref{BMPVSCHM},\eqref{BMPVROTM} and \eqref{BMPVSU2A}-\eqref{BMPVSU2M} . Next we computed the $\log \, T$ correction to the near-extremally extended BMPV solution. The results are given in equations \eqref{BMPVSCH},\eqref{BMPVROT} and \eqref{BMPVSU2}.
\end{itemize}

Next we discuss some important aspects of our results. Firstly, we would like to comment on various near-extremal limits in the context of the BTZ black holes. Under a small temperature expansion, the two horizons of the BTZ solution are separated by the small parameter $\epsilon = \frac{\ell^2}{r_0\beta}$. Just like we treated the length scales $\ell$ and $r_0$ to be independent in case of the extremal solution, one can naively attempt to introduce three independent length scales $\ell,r_0,\beta$ in case of the near-extremal solution. Now, various distinct hierarchy of scales can be considered that render the separation of two horizons small and declare them as distinct near-extremal limits. In these distinct limits, one can analyze the modular-invariant CFT$_2$ partition function and project the result to the vacuum character following \cite{Ghosh:2019qsp,Aggarwal:2023peg}. This would generate distinct universal logarithmic contributions to the logarithm of partition function. However, when we try to recover these distinct limits from the gravitational computation, it turns out we do not have the freedom to choose such distinct limits in our current formalism. The gravitational path integral on the near-extremal background can only be computed perturbatively with perturbation parameter $\epsilon\ll 1$, generating only logarithm of $\epsilon$. This is an interesting limitation in our perturbative computations. 

One important issue for a Euclidean path integral with a purely quadratic action is that the functional integral is Gaussian, and its convergence depends crucially on the sign of the quadratic form evaluated around a given gravitational saddle. Schematically, if
$$
Z\sim \int \mathcal D\phi\, e^{-S_E^{(2)}[\phi]},
$$
then fluctuation modes for which the quadratic action has the conventional positive sign may be integrated along the real contour, giving a convergent Gaussian. By contrast, when a saddle gives the opposite sign for a quadratic mode, integration along the real direction is divergent, and the corresponding contour must be deformed into an appropriate complex, often purely imaginary, direction so that the real part of the exponent becomes negative. Such contour deformations are closely related to the admissibility of complex metrics advocated by Kontsevich and Segal and subsequently applied to gravitational path integrals by Witten. More generally, the necessity of choosing nontrivial integration contours for quadratic fluctuations in Euclidean gravity, and the relation between the spectrum of the quadratic fluctuation operator and the appropriate contour, have been studied in detail in \cite{Kontsevich:2021dmb,Witten:2021nzp,Marolf:2022ntb,Liu:2023jvm}%{KWS}
. For the rotational zero modes of the BMPV solution, we encounter similar situation. Different values of the rotation parameter $\omega$ gives us different saddles and depending on the range of the rotation parameter $\omega$, we found the quadratic action changes its sign. Thus, to make sense of the corresponding 1-loop path integral, we had to choose the contour appropriately. Finally, with appropriate choice of the complex steepest-descent contour, we get finite contribution to the $\log \, T$ term.

We end the paper with a list of open problems: 
\begin{enumerate}
    \item 
The minimal BMPV black hole that we have studied here is a solution of five-dimensional $\mathcal{N}=2$ minimal supergravity. It is a $\frac{1}{2}$-BPS solution and therefore preserves four of the eight real supercharges.The standard BMPV background is a purely bosonic classical solution. The non vanishing bosonic fields are the metric $g_{\mu\nu}$ and the graviphoton gauge field $A_\mu$, while the fermionic fields are set to zero,$\psi_\mu=0.$ The fermionic completion of the BMPV black hole is obtained by acting on the bosonic BMPV solution with the broken supersymmetry transformations of five-dimensional $\mathcal{N}=2$ minimal supergravity.

Since the BMPV solution is $\frac{1}{2}$-BPS, it preserves four of the eight real supercharges and therefore has four broken real supercharges. These broken supersymmetries generate four fermionic zero modes. The complete configuration is of the form
\begin{equation}
\left(g_{\mu\nu},A_\mu,\psi_\mu\right),\nonumber
\end{equation}
where the gravitino $\psi_\mu$ is non vanishing.
Starting from the purely bosonic BMPV background, $
\psi_\mu^{(0)}=0,$
the leading fermionic field is generated by a supersymmetry transformation,
\begin{equation}
\psi_\mu^{(1)}
= \delta_\epsilon\psi_\mu
=
\left[
\nabla_\mu
+
\frac{1}{4\sqrt{3}}
\left(
\gamma_{\mu\nu\rho}
- 4g_{\mu\nu}\gamma_\rho
\right)
F^{\nu\rho}
\right]\epsilon, \nonumber
\end{equation}
where $\epsilon$ belongs to the broken-supersymmetry sector. Successive supersymmetry transformations generate corrections to both the fermionic and bosonic fields. Schematically,
\begin{align}
\begin{split}
\psi_\mu
&=
\psi_\mu^{(1)}
+
\psi_\mu^{(3)}
+\cdots, \nonumber \\
g_{\mu\nu}
&=
g_{\mu\nu}^{\rm BMPV}
+
g_{\mu\nu}^{(2)}
+
g_{\mu\nu}^{(4)}
+\cdots, \nonumber\\
A_\mu
&=
A_\mu^{\rm BMPV}
+
A_\mu^{(2)}
+
A_\mu^{(4)}
+\cdots. \nonumber
\end{split}
\end{align}

Since the supersymmetry parameters are Grassmann-valued, the expansion terminates after a finite number of terms. The resulting configuration is commonly referred to as the fermionic superpartner, fermionic wig, or fermionic completion of the BMPV solution. It would be good to study the $\log \, T$ contributions for this system. 
\item 
In \cite{Alvarado:2026kio} , the authors have studied the effect of higher derivative gravity interactions on the $\log \, T$ term, it would be nice to see the effects of rotation for such systems. In presence of rotation one should consider \cite{Banerjee:2013gvy} and perform the analysis. We shall report on this in the near future. 
\item The five dimensional rotational systems are much more generalized than the one we have studied in the present paper. In the corresponding gauged version of the theory, they are also asymptotically AdS, thus provides a larger parameter space to study the quantum corrected entropy. It would be interesting to extend the present analysis for to the entire parameter space.
\item In \cite{Liu:2017vll}%{1707.04197}
, \cite{Liu:2017vbl} %{1711.01076}
and subsequent papers \cite{PandoZayas:2020iqr,Hristov:2021zai,David:2021eoq,Bobev:2023dwx}%{2008.03239, 2107.12398,2112.09444,HEP 04 (2024) 020}
, authors suggest that, for a class of asymptotically AdS$_4$ black holes, there is an apparent issue with the zero modes contribution from the near -horizon and complete geometry. They concluded that the near-horizon computations do not match with the microscopic results. This seems to suggest that for asymptotically AdS black holes, the quantum entropy functions need appropriate modifications in incorporating the zero modes contributions. With our refined way of constructions of the zero modes, it would be interesting to make this issue clarified. Further, in a series of papers \cite{Dutta:2026zrt,Dutta:2025ypr,Dutta:2025uch}, the authors have constructed a microscopic description for three dimensional black flower states. It would be interesting to check if the quantum log corrections can reproduce the same.

\end{enumerate}

\acknowledgments{
We thank Ashoke Sen and Diksha Jain for  discussions on the project. We also thank Suvankar Dutta for inputs in simplifying mathematica codes. We also acknowledge the use of AI in improving the writeup of the manuscript. KG would like to thank IISER Pune Gravity workshop for hospitality where part of this work was presented. The work of NB is supported by SERB POWER fellowship SPG/2022/000370. NB would also like to thank ICTP regular associateship program. KG (UGC-Ref. No.: 231610142154) is supported by the University Grants Commission (UGC), New Delhi. The work of MS is supported by Department of Atomic Energy, Government
of India, under project no. RTI4019. We thank the people of India for their generous support to the advancement of basic sciences.}

\appendix
\section{Near-Horizon BMPV Geometry and the Kerr/CFT Form}\label{app:Kerr CFT form}
In the literature one frequently encounters two equivalent representations of the near-horizon BMPV geometry. The first is written in terms of the left-invariant one-forms on
$SU(2)\simeq S^3$. The second is the canonical near-horizon form commonly used in Kerr/CFT analyses.
The purpose of this section is to derive the second form from the first.
%\subsection{Near-horizon BMPV geometry in terms of left-invariant one-forms}
\subsection{Hopf coordinates to SU(2) left-invariant one-forms}
The near-horizon BMPV metric in Hopf coordinates \eqref{NH_BMPV_Hopf} is given by,
\begin{equation}
ds^2=
-\left(
q\rho d\tau
+
\frac{J}{8q^2}\tilde{\omega}
\right)^2
+
q^2\frac{d\rho^2}{\rho^2}
+
4q^2 d\Omega_3^2,
\end{equation}
with
\begin{equation}
d\Omega_3^2
=
d\tilde\theta^2
+
\sin^2\tilde\theta\, d\varphi^2
+
\cos^2\tilde\theta\, d\tilde\psi^2.
\end{equation}
% \begin{equation}
% Q=4q^2.
% \end{equation}
The Hopf coordinates are
\begin{equation}
0\le \tilde\theta \le \frac{\pi}{2},
\qquad
0\le \varphi<2\pi,
\qquad
0\le \tilde\psi<2\pi,
\end{equation}
and the one-form appearing in the BMPV solution is
\begin{equation}
\tilde{\omega}
=
\sin^2\tilde\theta\, d\varphi
-
\cos^2\tilde\theta\, d\tilde\psi.
\end{equation}
Introduce the Euler angles
\begin{equation}
\hat\theta=2\,\tilde\theta,
\qquad
\phi=\varphi+\tilde\psi,
\qquad
\chi=\tilde\psi-\varphi,
\end{equation}
with
\begin{equation}
0\le \hat\theta \le \pi,
\qquad
0\le \phi <2\pi,
\qquad
0\le \chi <4\pi,
\end{equation}
and the following set of left-invariant one-forms,
\begin{align}
\sigma_1 &=
\cos\chi\, d\hat\theta
+
\sin\chi \sin\hat\theta\, d\phi,
\\[5pt]
\sigma_2 &=
-\sin\chi\, d\hat\theta
+
\cos\chi \sin\hat\theta\, d\phi,
\\[5pt]
\sigma_3 &=
d\chi+\cos\hat\theta\, d\phi,
\end{align}
satisfying,
\begin{equation}
d\sigma_i
=
-\frac12\epsilon_{ijk}\,\sigma_j\wedge\sigma_k,\quad d\Omega^2_3=\frac14\sum_{i=1}^3\,\sigma_i^2,\quad \tilde\omega=-\frac12\sigma_3.
\end{equation}
In terms of Euler angles and the $SU(2)$ left-invariant one-forms introduced above, the near-horizon BMPV metric takes the following form,
\begin{equation}
ds^2
=
\frac Q4\left[
-(r\,dt+\tilde J \sigma_3)^2
+\frac{dr^2}{r^2}
+\sigma_1^2+\sigma_2^2+\sigma_3^2
\right],
\end{equation}
where,
\begin{equation}
r=\rho,\quad t=\tau,\quad \tilde J=\frac{J}{2Q^{3/2}}.
\end{equation}
Since
\begin{equation}
\sigma_1^2+\sigma_2^2
=
d\hat\theta^2
+
\sin^2\hat\theta\, d\phi^2,
\end{equation}
we substitute the same back into the metric to get,
\begin{align}
    ds^2&=\frac{Q}{4}\,\bigg[-r^2dt^2+\frac{dr^2}{r^2}+d\hat\theta^2+\sin^2\hat\theta\,d\phi^2+(1-\tilde J^2)\sigma_3^2-2\,\tilde J\, r\,dt\,\sigma_3\bigg]
\end{align}
\subsection{Re-arrangement and time re-scaling}
The terms involving $\sigma_3$ can be re-arranged as shown,
\begin{align}
(1-\tilde J^2)\sigma_3^2
-
2\tilde J r\,dt\,\sigma_3
&=(1-\tilde J^2)
\bigg(
\sigma_3
-
\frac{\tilde J}{1-\tilde J^2}
\,r\,dt
\bigg)^2-
\frac{\tilde J^2}{1-\tilde J^2}
\,r^2dt^2.
\end{align}
Substituting this identity into the metric yields
\begin{align}
ds^2
=
\frac{Q}{4}\Bigg[
&
-\frac{r^2}{1-\tilde J^2}\,dt^2
+
\frac{dr^2}{r^2}
+
d\hat\theta^2
+
\sin^2\hat\theta\, d\phi^2
+
(1-\tilde J^2)
\bigg(
\sigma_3
-
\frac{\tilde J}{1-\tilde J^2}
\,r\,dt
\bigg)^2
\,\Bigg]
\end{align}
Under a time re-scaling given by,
\begin{equation}
\hat t
=
\frac{t}{\sqrt{1-\tilde J^2}}.
\end{equation}
the metric becomes
\begin{equation}
ds^2
=
\frac{Q}{4}
\left[
-r^2d\hat t^2
+
\frac{dr^2}{r^2}
+
d\hat\theta^2
+
\sin^2\hat\theta\, d\phi^2
+
(1-\tilde J^2)
\left(
\sigma_3
-
k\,r\,d\hat t
\right)^2
\right].
\end{equation}
where,
\begin{equation}
    k=\frac{\tilde J}{\sqrt{1-\tilde J^2}}.
\end{equation}
\subsection{To Kerr/CFT form}
To re-write the metric in Euclidean signature we perform a Wick rotation,
\begin{equation}
    \hat t\rightarrow-i\tau_E
\end{equation}
The metric in Euclidean signature is given by,
\begin{equation}
ds^2
=
\frac{Q}{4}
\left[
r^2d\tau_E^2
+
\frac{dr^2}{r^2}
+
d\hat\theta^2
+
\sin^2\hat\theta\, d\phi^2
+
(1-\tilde J^2)
\left(
\sigma_3
+
k_E\,r\,d\tau_E
\right)^2
\right].\label{eq:metric1}
\end{equation}
where,
\begin{equation}
    k_E=i\,k.
\end{equation}
Introduce the $AdS_2$-Rindler coordinates $(\theta,\eta)$,
\begin{equation}
r
=
\cosh\eta+\sinh\eta\,\cos\theta ,
\qquad\tau_E
=
\frac{
\sinh\eta\,\sin\theta
}
{
\cosh\eta+\sinh\eta\,\cos\theta
}.
\end{equation}
As a result, the fiber term in \eqref{eq:metric1} becomes,
\begin{equation}
d\chi+\cos\hat\theta\,d\phi
+k_E(\cosh\eta-1)d\theta
+k_E\,d\lambda.
\end{equation}
Define a new fiber coordinate
\begin{equation}
\psi
=
\chi+k_E\lambda,
\end{equation}
where,
\begin{equation}
\lambda
=
\theta
-
2\tan^{-1}
\left[
\frac{
\tan(\theta/2)
}
{
\cosh\eta+\sinh\eta
}
\right].
\end{equation}
and now, the fiber term is given by,
\begin{equation}
d\psi+\cos\hat\theta\,d\phi
+k_E(\cosh\eta-1)d\theta .
\end{equation}
Under the above change of near-horizon coordinates followed by the substitution,
\begin{equation}
1-\tilde J^2
=
\cosh^2\omega
\end{equation}
the near-horizon BMPV metric in Euclidean signature takes the following form,
\begin{equation}
ds^2
=
\frac{Q}{4}
\left[
\sinh^2\eta \,d\theta^2
+d\eta^2
+
d\theta^2
+
\sin^2\hat\theta\, d\phi^2
+
\cosh^2\omega
\left(
d\psi+\cos\hat\theta\,d\phi
-\tanh\omega\,(\cosh\eta-1)\,d\theta
\right)^2
\right],
\end{equation}
where, the $AdS_2$ factor is now explicit. Changing $\omega\rightarrow i\omega$ takes us to the Euclidean, near-horizon metric of \cite{Gomes:2013cca} while, $\omega\rightarrow-i\omega$ would take us to the form in \eqref{BMPV_Kerr-CFT} which also aligns with the form given in \cite{Kolanowski:2024zrq},
\begin{equation}
    ds^2=
\frac{Q}{4}
\left[
\sinh^2\eta\,d\theta^2
+d\eta^2
+
d\hat\theta^2
+
\sin^2\hat\theta\, d\phi^2
+
\cos^2\omega
\left(
d\psi+\cos\hat\theta\,d\phi+i\tan\omega(\cosh\eta-1)d\theta
\right)^2
\right].
\end{equation}
\section{Lie dragging of gauge fields}\label{Lie_drag_U(1)}%{\bf send to appendix}\\
Interestingly, the background gauge field $\bar A_\mu(x)$ is also Lie dragged along each vector field $\xi^\mu(x)$ generating the above diffeomorphisms (Schwarzian modes, rotational modes, and $SU(2)$ vector modes). The gauge fluctuation $a_\mu(x)$ so produced takes the following form in general,
\begin{align}
    a_\mu(x)&=\mathcal{L}_\xi \bar A_\mu=\xi^\nu \bar \nabla_\nu \bar A_\mu+\bar A_\nu  \bar \nabla_\mu \xi^\nu
\end{align}
Apriori there is no reason for such fluctuations to remain on the gauge slice and hence, be considered physical. In case, they are not we must transform them by a compensating gauge transformation $ \bar \nabla_\mu \Lambda$ such that the gauge-fixing condition (governing the chosen gauge slice) is satisfied which in our case is,
\begin{align}
     \bar \nabla^\mu a_\mu&=0
\end{align}
The gauge transformed fluctuation $a'_\mu(x)$ is given by,
\begin{align}
    a_\mu\longrightarrow a'_\mu=a_\mu+ \bar \nabla_\mu\Lambda
\end{align}
Additionally, for some cases (viz., the Schwarzian and the rotational) one may find a compensating gauge parameter $\Lambda$ such that $a'_\mu=$ const. and subsequently, $ \bar \nabla^\mu a'_\mu=0$. Thus, metric and gauge fluctuations are decoupled. Equivalently, the above demand corresponds to,
\begin{align}
    \mathcal{L}_\xi\,\bar F_{\mu\nu}&=0
\end{align}
The proof is shown below. Before we begin with the same, it is instructive to re-write the Lie drag of the background gauge field as follows,
\begin{align}
    \mathcal{L}_\xi\bar A_\mu&=-\,\xi^\nu\bar F_{\mu\nu}\,+\, \bar \nabla_\mu(\bar A\cdot\xi)
\end{align}
Without loss of generality we set,
\begin{align}
    &\quad a'_\mu=0\Rightarrow  \bar \nabla_\mu a'_\nu- \bar \nabla_\nu a'_\mu=0,\\[5pt]
    &\Rightarrow  \bar \nabla_\mu(-\,\xi^\rho\bar F_{\nu\rho}\,+\, \bar \nabla_\nu(\bar A\cdot\xi)+ \bar \nabla_\nu\Lambda)\,-\,(\mu\leftrightarrow\nu)=0,\\[5pt]
    &\Rightarrow \bar F_{\rho\nu} \bar \nabla_\mu\xi^{\rho}+\bar F_{\mu\rho} \bar \nabla_\nu\xi^{\rho}+\xi^\rho( \bar \nabla_\nu\bar F_{\mu\rho}+ \bar \nabla_\mu\bar F_{\rho\nu})+\underbrace{[ \bar \nabla_\mu, \bar \nabla_\nu]\overbrace{(\bar A\cdot\xi+\Lambda)}^{\rm scalar}}_{=0}=0,\\[5pt]%\qquad\big(\,[D_\mu,D_\nu]\phi=0\,\big)
    &\Rightarrow \bar F_{\rho\nu} \bar \nabla_\mu\xi^{\rho}+\bar F_{\mu\rho} \bar \nabla_\nu\xi^{\rho}+\xi^\rho  \bar \nabla_\rho\bar F_{\mu\nu}=0,\qquad(\text{Bianchi Identity})\\[5pt]
    &\Rightarrow \mathcal{L}_\xi\,\bar{F}_{\mu\nu}=0.
\end{align}
Thus, we solve for $\Lambda(x)$ from,
\begin{align}
    a'_\mu=-\,\xi^\nu\bar F_{\mu\nu}\,+\, \bar \nabla_\mu(\bar A\cdot\xi+\Lambda)=0.
\end{align}

\section{First and second variations around an on-shell background}\label{APPC}

Consider the gauge fixed action:
\begin{equation}
S_{\rm tot}
=
S_{\rm EMCS}
+
S_{\rm gf}
+
S_{\rm gf}.
\end{equation}
The three components arising above are defined in the main draft. In particular $S_{\rm EMCS}$ is the gauge invariant part of the action. The displacement from the background is
\begin{equation}
\delta g_{\mu\nu}=h_{\mu\nu},
\qquad
\delta A_\mu=a_\mu.
\end{equation}
These are generic fluctuations (not necessarily gauge transformations). The first variation evaluated on the background $(\bar g, \bar A)$ is
\begin{equation}
\delta S_{\rm inv}
=\frac{1}{16\pi G_5}\int d^5x\,
\left(
\bar E_{g}^{\mu\nu}h_{\mu\nu}
+\bar E_{A}^\mu a_\mu
\right)=0.
\end{equation}

Here we already have incorporated the Gibbons-Hawking-York terms, so all boundary terms are set to zero. The variations of the fluctuations are also zero at the boundary (this is true when we wish to extremize any action for finding equation of motion). 
Next we find the linearized equations around the background:
\begin{align}
\delta E_{\mu\nu}[h,a]
&=E_{\mu\nu}[h]+E_{\mu\nu}[a],\\
\delta E_A^\mu[h,a]
&=E_A^\mu[h]+E_A^\mu[a].
\end{align}
Then, up to boundary terms, the gauge-invariant quadratic action is schematically
\begin{equation}
S_{\rm ECMS}^{(2)}
=\frac{1}{32\pi G_5}\int d^5x\,\
\left[
 h_{\mu\nu}\,\delta E^{\mu\nu}[h,a]
+a_\mu\,\delta E_A^\mu[h,a]
\right].
\end{equation}

Now, let us specialize to a case, where the field fluctuations are just gauge (diffeo and U(1)).Let \(\xi^\mu\) generate an infinitesimal diffeomorphism. The metric fluctuation is
\begin{equation}
h_{\mu\nu}=\mathcal{L}_\xi \bar g_{\mu\nu}
=\bar \nabla_\mu\xi_\nu+\bar \nabla_\nu\xi_\mu.
\end{equation}
Including the U(1) transformation for the gauge field, the fluctuation is
\begin{align}
a_\mu
&=\mathcal{L}_\xi \bar A_\mu+\bar \nabla_\mu\Lambda\\
&=\xi^\nu \bar F_{\nu\mu}
+\bar \nabla_\mu\left(\xi^\nu \bar A_\nu+\Lambda\right).
\end{align}
It is convenient to define
\begin{equation}
\lambda\equiv \xi^\nu \bar A_\nu+\Lambda,
\end{equation}
so that
\begin{equation}
a_\mu=\xi^\nu \bar F_{\nu\mu}+\bar \nabla_\mu\lambda.
\end{equation}
The distinction between \(\Lambda\) and \(\lambda\) is useful: \(\Lambda\) is the additional \(U(1)\) gauge parameter accompanying the diffeomorphism, while \(\lambda\) is the total exact piece that natually arises if one uses Cartan's identity.
Now the diffeomorphism covariance of the equations of motions gives
\begin{equation}
\delta E_{\mu\nu}^{\rm inv}
[\mathcal{L}_\xi \bar g,\mathcal{L}_\xi \bar A+d\Lambda]
=\mathcal{L}_\xi \bar E_{\mu\nu}=\xi^\rho \bar \nabla_\rho \bar E_{\mu\nu}
+\bar E_{\rho\nu}\bar \nabla_\mu\xi^\rho
+\bar E_{\mu\rho}\bar \nabla_\nu\xi^\rho,.
\end{equation}
The compensating \(d\Lambda\) does not appear because the gauge-invariant Einstein equation is \(U(1)\)-gauge invariant.

Similarly,
\begin{equation}
\delta E_A^{\mu,{\rm inv}}
[\mathcal{L}_\xi \bar g,\mathcal{L}_\xi \bar A+d\Lambda]
=\mathcal{L}_\xi \bar E_A^{\mu}=\xi^\rho\bar \nabla_\rho \bar E_A^{\mu}
-\bar E_A^{\rho}\bar \nabla_\rho\xi^\mu.
\end{equation}
Again there is no explicit \(\Lambda\) or \(\lambda\): the gauge-invariant Maxwell--Chern--Simons equation depends on the gauge potential only through gauge-invariant combinations (up to the standard boundary behavior of the Chern--Simons action), and the local Euler--Lagrange equation is Abelian gauge invariant. Further since the background is on shell,
\begin{equation}
\delta E_{\mu\nu}^{\rm inv}=0,
\qquad
\delta E_A^{\mu,{\rm inv}}=0.
\end{equation}
Thus every infinitesimal combined diffeomorphism plus Abelian gauge transformation is a null direction of the \emph{gauge-invariant} Hessian, modulo boundary terms. 

However there is the crucial distinction, when the gauge fixed terms are included in the action. The equations of the gauge-fixed quadratic theory contain the gauge-fixing operators in addition to the gauge-invariant linearized equations.

For the harmonic functional
\begin{equation}
\mathcal{F}_\mu=\bar \nabla^\nu h_{\mu\nu}-\frac12\bar \nabla_\mu h,
\end{equation}
variation of the quadratic gauge-fixing action gives, up to the common normalization convention,
\begin{equation}
(\delta E_{\mu\nu})_{\rm gf}
=\bar \nabla_{(\mu}\mathcal{F}_{\nu)}
-\frac12 \bar g_{\mu\nu}\bar \nabla_\rho\mathcal{F}^\rho.
\end{equation}
Defining,
\begin{equation}
\mathcal K_{\mu\nu}{}^\rho
=\delta^\rho{}_{(\nu}\bar \nabla_{\mu)}
-\frac12 \bar g_{\mu\nu}\bar \nabla^\rho, \quad
\text{we have} \quad
(\delta E_{\mu\nu})_{\rm gf}
=\mathcal K_{\mu\nu}{}^\rho\mathcal{F}_\rho.
\end{equation}

For the Lorenz functional $
\mathcal{G}=\bar \nabla^\mu a_\mu,
$ 
variation of the Maxwell gauge-fixing term gives
\begin{equation}
(\delta E_A^\mu)_{\rm gf}=\bar \nabla^\mu\mathcal{G}. \quad \text {and hence}
\quad
\mathcal K_A^\mu=\bar \nabla^\mu.
\end{equation}

Accordingly the full gauge-fixed linearized equations may be written schematically as
\begin{equation}
\delta E_{\mu\nu}^{\rm GF}
=\delta E_{\mu\nu}^{\rm inv}
+\mathcal K_{\mu\nu}{}^\rho\mathcal{F}_\rho[h],
\end{equation}
\begin{equation}
\delta E_A^{\mu,{\rm GF}}
=\delta E_A^{\mu,{\rm inv}}
+\mathcal K_A^\mu\mathcal{G}[a].
\end{equation}
The symbols ``GF'' here denote the equations of the gauge-fixed \emph{quadratic fluctuation theory}; they should not be confused with the classical background equations, for which the quadratic gauge-fixing functionals vanish at zero fluctuation. For the special fluctuations that is generated by a pure gauge transformation of the background $(\bar g, \bar A)$, we have
\begin{equation}
\mathcal{F}_\mu
=(\bar \Box\delta_\mu{}^\nu+\bar R_\mu{}^\nu)\xi_\nu, \qquad
\mathcal{G}
=\bar \nabla^\mu(\xi^\nu \bar F_{\nu\mu})+\bar \Box\lambda.
\end{equation}
Further, if these fluctuations satisfy residual gauge fixing condition, then ,
\begin{equation}
(\bar \Box\delta_\mu{}^\nu+\bar R_\mu{}^\nu)\xi_\nu=\bar \nabla^\mu(\xi^\nu \bar F_{\nu\mu})+\bar \Box\lambda=0.
\end{equation}
Therefore, we see that when the pure gauge fluctuations satisfy residual gauge fixing condition,
\begin{align}
\delta E_{\mu\nu}^{\rm GF}
&=\mathcal{L}_\xi \bar E_{\mu\nu}
+\mathcal K_{\mu\nu}{}^\rho
(\bar \Box\delta_\rho{}^\sigma+\bar R_\rho{}^\sigma)\xi_\sigma=0,
\\
\delta E^{\mu,{\rm GF}}
&=\mathcal{L}_\xi \bar E_A^{\mu}
+K_A^\mu\left[
\bar \nabla^\rho(\xi^\nu \bar F_{\nu\rho})+\bar \Box\lambda
\right]=0.
\end{align}
This makes the role of \(\lambda\) completely explicit. It does \emph{not} enter the gauge-invariant Lie-drag identities. It enters only through the Lorenz gauge-fixing functional, and is chosen precisely so that this extra gauge-fixing contribution vanishes. Finally, for these special fluctuations, the quadratic action vanishes, i.e.

\begin{equation}
S_{\rm GF}^{(2)}
=\frac{1}{32\pi G_5}\int d^5x\,
\left[
 h_{\mu\nu}\,\delta E_{GF}^{\mu\nu}[h,a]
+a_\mu\,\delta E_{GF}^{A \mu}[h,a]
\right]=0.
\end{equation}

The above equations clearly shows that once the residual gauge conditions are satisfied, the fluctuations are the zero modes of the corresponding kinetic operators in the background.

\bibliography{near-BMPV}

\end{document}